\documentclass[10pt,journal,compsoc]{IEEEtran}
\ifCLASSOPTIONcompsoc
\else
\usepackage{cite}
\fi

\usepackage{times}
\usepackage{soul}
\usepackage{url}
\usepackage{graphicx}
\usepackage{amsmath}
\usepackage{amsthm}
\usepackage{booktabs}
\usepackage[switch]{lineno}
\usepackage{makecell}
\usepackage{color}
\usepackage{graphicx}
\usepackage{framed}
\usepackage{color}
\usepackage{amsmath,amssymb}
\usepackage{setspace}
\usepackage{makecell}
\usepackage{amsthm}
\usepackage{color}
\usepackage{subcaption}
\usepackage{float}
\usepackage{enumitem}
\usepackage{subfloat}
\usepackage{pict2e}
\usepackage{circledsteps}
\usepackage{enumitem}
\usepackage[ruled,vlined,linesnumbered,ruled]{algorithm2e}
\usepackage{cite}
\usepackage{amsmath,amssymb,amsfonts}
\usepackage{graphicx}
\usepackage{textcomp}
\usepackage{xcolor}
\usepackage{multirow}
\usepackage{color}
\usepackage{amsmath,amssymb}
\usepackage{setspace}
\usepackage{makecell}
\usepackage{amsthm}
\usepackage{color}
\usepackage{booktabs}
\usepackage{balance}
\usepackage{subcaption}
\usepackage{array}
\usepackage{flushend}
\usepackage{balance}
\usepackage{url}
\usepackage{ragged2e}
\renewcommand{\justify}{\leftskip=0pt \rightskip=0pt plus 0cm}

\newtheorem{myDef}{Definition}

\newcommand{\bcircled}[1]{\Circled[fill color=black, inner color=white]{#1}}
\newlist{frameworklist}{itemize}{1}
\setlist[frameworklist]{label=-}

\newcommand{\zqy}[1]{{\color{blue}#1}}

\newtheorem{example}{Example}[section]
\newtheorem{theorem}{Theorem}[section]

\title{Flexible Keyword-Aware Top-$k$ Route Search 
}
\begin{document}
\author{\IEEEauthorblockN{{Ziqiang Yu}$^1$, Xiaohui Yu$^{2*}$\thanks{*Corresponding author}, Yueting Chen$^3$, Wei Liu$^1$, Anbang Song$^1$, Bolong Zheng$^4$} \\
\IEEEauthorblockA{\small
$^1$Yantai University 
\hspace{0.2em}$^2$York University
\hspace{0.2em}$^3$Seattle University
\hspace{0.2em}$^4$Wuhan University of Technology
}
}

\IEEEtitleabstractindextext{%
\begin{abstract}
\justify{
With the rise of Large Language Models (LLMs), tourists increasingly use it for route planning by entering keywords for attractions, instead of relying on traditional manual map services. LLMs provide generally reasonable suggestions, but often fail to generate optimal plans that account for detailed user requirements, given the vast number of potential POIs and possible routes based on POI combinations within a real-world road network. In this case, a route-planning API could serve as an external tool, accepting a sequence of keywords and returning the top-$k$ best routes tailored to user requests. To address this need, this paper introduces the Keyword-Aware Top-$k$ Routes (KATR) query that provides a more flexible and comprehensive semantic to route planning that caters to various user's preferences including flexible POI visiting order, flexible travel distance budget, and personalized POI ratings. Subsequently, we propose an explore-and-bound paradigm to efficiently process KATR queries by eliminating redundant candidates based on estimated score bounds from global to local levels. Extensive experiments demonstrate our approach's superior performance over existing methods across different scenarios.
}
\end{abstract}

}

\maketitle

\section{Introduction}

Keyword-based route planning has been extensively studied and implemented in real-world systems for decades~\cite{kanza2010interactive,rice2013engineering,li2012optimal,Shang2012User,6816646,chan2021cost,li2021optimal,costa2015optimal,tong2016online,Shang2020survey,li2021towards,li2021traffic, li2005trip,roy2011interactive,hashem2013group,dai2016personalized,zhu2022top,luo2023task,cao2012keyword,rice2013parameterized,bao2013efficient,zhang2012multi,zeng2015optimal,lu2016efficient,liang2018top,liu2018finding,chen2008multi,app112110497,tong2022hu,li2023finding,10.1145/2666310.2666411}. These systems typically require users to input specific keywords to identify points of interest (POIs) and then generate routes based on the identified POIs. 
For instance, consider {a user traveling through an unfamiliar city who intends to visit two prominent landmarks but also desire the flexibility to grab coffee}, enjoy local cuisine for lunch and dinner, and refuel their car as needed throughout the day. In such cases, certain keywords (e.g., landmark names) might map to specific POIs, while others (e.g., “coffee,” “local food,” or “refuel”) are more generic and correspond to multiple possible POIs. Importantly, the sequence in which these activities are undertaken may not be important, as long as they are all completed within the route.



{As systems powered by Large Language Models (LLMs) like ChatGPT~\cite{zhao2023survey} become more prevalent, users increasingly want to use them for tourist route planning via keywords or conversational inputs. However, due to inherent limitations in LLMs \cite{zhang2023siren, gao2023retrieval}, they may offer generally reasonable suggestions but often cannot generate optimal route plans that meet detailed user requirements, given the vast number of POIs and possible routes a real-world road network. A common approach to overcoming these limitations involves creating an agent that integrates LLMs with external tools \cite{ruan2023tptu, hsieh2023tool}. 
In our work, we investigate an efficient route-planning algorithm that could serve as such a tool, accepting a sequence of keywords and returning the top-$k$ best routes tailored to user preferences. Once LLMs receive a query, they will extract keywords from user conversations and tell the agent to invoke this search algorithm to provide high-quality, personalized route plans.

However, the context of LLM-based route planning introduces new requirements for the underlying algorithms. First, to ensure a user-friendly experience, the system should avoid asking users to input complex constraints, beyond specifying a starting point and a set of keywords indicating the desired POIs, to keep the query process simple. Second, in a multi-user system, individual preferences may vary significantly, both in terms of POI selection and route characteristics. For example, one user might prioritize visiting highly rated POIs, while another might focus on minimizing travel distances. Therefore, even with minimal input, the algorithm must support a more flexible, comprehensive approach that accounts for these diverse factors. The key features necessary for such an algorithm are outlined below:
}

{\bf R1: Flexible Visiting Order.} The order of visiting POIs along the route should not rigidly follow the sequence provided in the keyword input. While a fixed order can be specified if necessary, it should remain optional, as users may prefer a shorter or more efficient route. For example, whether refueling the car occurs before or after lunch is less important as long as it is convenient within the overall route.

{\bf R2: Flexible Distance Budget.} 
Travel distance should not be the sole constraint in route planning. Some users may prioritize visiting more appealing or unique POIs, even if they require traveling farther, over adhering to strict distance limitations.

{\bf R3: Personalized POIs Ratings.} Users often evaluate POIs based on factors such as popularity, cost, or alignment with their preferences and keywords. The route planning process must incorporate these diverse rating criteria to tailor recommendations to individual user preferences effectively.

\zqy{

} 

Such keyword-based route planning problems have been extensively studied, but existing solutions often fail to meet the above requirements due to rigid constraints. One approach ~\cite{kanza2010interactive, rice2013engineering, li2012optimal, chan2021cost, li2021optimal, costa2015optimal, li2021towards, li2021traffic, li2005trip, roy2011interactive, hashem2013group,dai2016personalized,zhu2022top} requires tourists to visit POIs in the exact input keyword order, leading to the Optimal Sequenced Routes (OSR) problem, but which cannot satisfy the requirement {\bf R1}. 
{Although algorithms addressing the keyword-aware optimal route (KOR) problem \cite{cao2012keyword, rice2013parameterized, bao2013efficient,zhang2012multi,zeng2015optimal,liang2018top,liu2018finding,chen2008multi,app112110497,li2023finding} relaxes the fixed order of POI visits, they impose strict travel distance budgets to factitiously restrict the search space. As the search scope expands, their query costs increase dramatically, failing to meet the requirement {\bf R2}. Moreover, these methods often ignore the varying ratings of POIs and treat all POIs with the same keyword equally to reduce complexity, but this fails to fulfill the requirement {\bf R3}.}

To address all the above requirements, we present KATR (Keyword-Aware Top-$k$ Routes), a flexible keyword-aware query semantic to search optimal routes by requiring only a starting point and relevant keywords. Given a road network with various POIs tagged with keywords and ratings, a KATR query $q(v_q, T_q)$ seeks to find $k$ optimal routes starting from a query point $v_q$ over the entire road network. These routes include POIs that cover all keywords of $T_q$ in any flexible order, without imposing travel distance budgets. The optimality of each route is determined by a scoring function that balances route distance against the cumulative POI ratings along the route. {Under this query paradigm, KATR emphasizes high query efficiency to meet the real-time interaction demands of LLM-based route planning, a goal shared by many existing route planning solutions.}

{Note that KATR can efficiently handle various scenarios and constraints often encountered in existing keyword-aware route planning solutions. (1) {\em Predefined Visiting Order:} If a specific order of keyword-tagged POIs is required, KATR focuses exclusively on this order, bypassing flexible visiting options, which significantly reduces computational complexity; (2) {\em Strict Distance Budgets:} If strict distance budgets are specified, KATR can readily use these budgets to limit the search space, simplifying the overall search process; (3) {\em Identical POI Rating:} In cases where all POIs have the same rating, KATR treats this as a special case and manages it easily without additional effort.} 

{While the KATR query semantic increases its applicability, it also creates an exponential number of candidates to evaluate.} 
{Thus, the primary challenge inherent in the KATR query is to effectively filter a large number redundant candidates by weighing both route distances and cumulative POI ratings, without losing the exact $k$ optimal routes. This challenge is not directly addressed by most existing OSR and KOR methods~\cite{cao2012keyword,lu2016efficient,liu2018finding,zhu2022top,li2023finding}, typically relying on the Progressive Neighbor Exploration (PNE) strategy~\cite{liu2018finding}.  
The main idea of PNE is to progressively explore neighboring km-POIs from the starting point, generating numerous partial candidate routes connecting the starting point to neighboring POIs, and then prioritize the expansion of the shorter ones until uncovering the shortest route covering all query keywords. However, applying this strategy to the KATR query presents several issues. First, the PNE principle generates numerous candidate partial routes in each iteration, making it particularly complex for KATR queries due to flexible POI visit orders that greatly increase the number of partial routes. Second, most PNE-based solutions focus on travel cost rather than POI rating, allowing for simple pruning based solely on route distance. However, this strategy struggles with the KATR query that considers both POI rating and route distance, as some remote but high-rated POIs will make this solution difficult to efficiently filter out irrelevant candidates and fail to early terminate the expansion.} 

{To address the above challenges, we propose KATR-Search that progressively narrows the search scope and stepwise approaches the top-$k$ optimal routes. The key insight is that while POI ratings might bring any POI into consideration, the actual optimal routes are likely within a limited area. Hence, KATR-Search first outlines this limited search space. It starts from the query vertex and performs heuristic exploration to find $k$ seed routes. These routes are used to initialize the limited search space smaller than the entire road network, yet guaranteed to contain the top-$k$ routes. The exploration proceeds within this space using a hierarchical pruning strategy, progressively narrowing the search as better routes are found, until finalizing the top-$k$ routes. Unlike PNE-based methods, which expand numerous candidates to gradually find better ones, KATR-Search assesses the maximum potential score of each candidate across different search space levels, facilitating early removal of less promising candidates based on route distance and POI ratings, while ensuring accuracy.
}

Our contributions are summarized as follows:

(1) We introduce KATR, a keyword-aware route search query specifically engineered to pinpoint the top-$k$ optimal routes. Unlike existing approaches, KATR operates without enforcing limitations on the sequence of POI visits, the scope of the search, or the ratings of POIs, offering unprecedented flexibility in route querying.



(2) Different from a PNE-Based solution that extends partial candidate routes locally, our approach introduces an explore-and-bound paradigm to filter  redundant candidates from global to local levels. This method prevents generating numerous irrelevant partial routes, thus reducing the number of routes requiring precise score calculation.

(3) We carry out comprehensive experiments on various real datasets, showcasing the effectiveness of KATR-Search compared to baseline methods across a range of scenarios.

The rest of this paper are organized as follows: 
Section~\ref{sec:relatd work} reviews related work on keyword-based route planning. Section~\ref{subsec:preliminary} defines the KATR query, and Section~\ref{sec:KATR-Search} introduces the KATR-Search solution. Section~\ref{sec:experiment} evaluates the solution's performance through experiments. Finally, Section~\ref{sec:conclusion} summarizes the paper.

\section{Related Work}\label{sec:relatd work}
Extensive research on keyword-based route planning can generally be categorized into two types: OSR queries and KOR queries, based on whether the POI visiting sequence is predetermined. 

{\bf OSR and its variants.} The OSR query has seen considerable exploration in literature~\cite{OSR-VLDB-2008,OSR-Geoinformatica-2008,cao2012keyword, Zheng2013Towards, liang2018top,liu2018finding,hao2019keyword,zhu2022top,tong2022unified}. In an OSR query, POIs are categorized by keyword tags, with the aim of finding the optimal route that passes through a specified sequence of POIs in the order dictated by keywords. One notable approach, Progressive Neighbor Exploration (PNE)~\cite{OSR-VLDB-2008}, addresses this by iteratively selecting the nearest relevant POI from the current location, constructing partial routes connecting the starting point to neighboring POIs, and extending the most promising partial route until all necessary POI categories are covered. StarKOSR~\cite{liang2018top} operates on a similar principle, utilizing 2-Hop-Labeling to expedite the search for the closest POI and extending this methodology to discover multiple optimal sequenced routes. ROSE-GM~\cite{zhu2022top} extends StarKOSR by incorporating a graph embedding method to locate the nearest POI. A challenge in these approaches is that the PNE principle may lead to a substantial number of candidate partial routes in each iteration, especially in scenarios without a predefined POI visiting order.

{\bf KOR and its variants.} The KOR query and its variants seek to find the shortest route from a source to a destination, passing through POIs in any sequence, as long as they cover all query keywords~\cite{cao2012keyword, rice2013parameterized, bao2013efficient,zhang2012multi,zeng2015optimal,lu2016efficient,liang2018top,liu2018finding,chen2008multi,10.1145/3292500.3330835,app112110497,li2023finding}. One common challenge is their tendency to impose strict constraints on the search scope to manage computational complexity. For instance, OSScaling\cite{cao2012keyword} directly implements a distance or time budget that halts the exploration of potential routes once they exceed this limit. Some approaches simplify the model by prioritizing geographically closer POIs within each category, ignoring the diversity in POI rating. Furthermore, many studies specify a destination, making it easier to exclude distant POIs from consideration. These measures often restrict the search scope, either directly or indirectly, to reduce computational complexity, without guaranteeing the preservation of the exact top-$k$ optimal routes~\cite{li2012optimal, dai2016personalized, IG-Tree-WWW-2019, li2023finding}. 

The intensive computation required by PNE techniques is amplified when applied to KOR queries due to the flexibility in POI visitation order, particularly for top-$k$ routes. DAPrune~\cite{li2023finding} addresses this by using path enumeration based on Yen's algorithm to find the top-$k$ routes covering all keywords. It starts by identifying the shortest initial path from the source to the destination, treating each vertex as a potential detour. The method then extends partial routes through these detours to neighboring POIs with unmet keywords, continuing until the route reaches the destination and includes all query keywords, forming candidate routes. Finally, it selects the $k$ shortest candidate routes as the results. However, this approach generates numerous partial routes during each extension, leading to exponential complexity as $k$ increases. Moreover, DAPrune's lack of consideration for varying POI ratings makes it unsuitable for KATR queries, which require diverse POI ratings for route scoring.{Moreover, some work~\cite{shang2017trajectory, Shang2012User,Zheng2013Towards, 10.1007/s00778-013-0331-0,shang2018parallel} use trajectory data to enhance keyword-based route planning, but they pay limited attention to relaxing constraints when recommending routes.}   

\section{Preliminaries and Problem Statement}\label{subsec:preliminary}

\begin{myDef}[Graph]\label{def:graph}
The road network is modeled as a weighted undirected graph $G=\left(V, E, W\right)$ such that 1) each vertex $v_i \in V$ represents an intersection in the road network, with a geographical coordinate (longitude, latitude); 2) each edge $e_{i,j}\in E$ denotes the road segment between intersections $v_i$ and $v_j$; and 3) any weight $w_{i,j} \in W$ denotes the length of the edge $e_{i,j}$.
\end{myDef}

\begin{figure}[htbp]
\centering
\vspace{-0.4cm}
\includegraphics[width=0.5\textwidth]{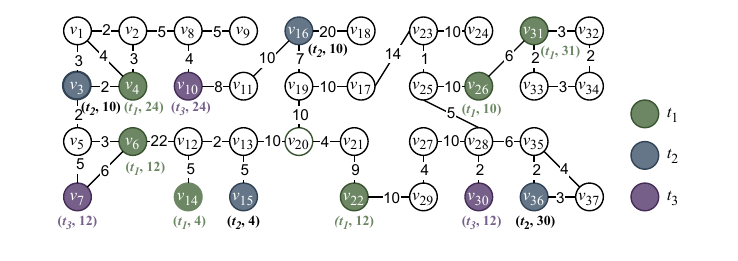}
 \vspace{-0.8cm}
\caption{Graph Instance with POIs}\label{fig:original-Graph}
\vspace{-0.3cm}
\end{figure}






{\bf POI.} In our problem setting, we position each POI at a vertex. Alongside the attributes of a regular vertex, a POI $v_i (t_i, \tau_i)$ incorporates an associated keyword $t_i$ and a rating $\tau_i$. For a POI lies along an edge $e_{i,j}$ connecting vertices $v_i$ and $v_j$, we establish a vertex corresponding to this POI, connecting $v_i$ and $v_j$. Thus, we use the same symbol to denote both POI and vertex. For simplicity, we assume that each vertex corresponds to only one POI in the subsequent discussion. Our approach also accommodates scenarios where multiple POIs locate at the same vertex. Moreover, we assume that both the POIs and query keywords come from the same domain, allowing for exact matches or mismatches exclusively. In our semantic, each POI is linked to a single keyword. In cases where a POI is associated with multiple keywords~\cite{lu2016efficient}, we place multiple POIs at the same location, each linked to a distinct keyword. A POI is considered km-POI if its keyword matches one from the query keywords.

The POIs on the road network shown in Figure~\ref{fig:original-Graph} have three different types of keyword tags $t_1$, $t_2$, and $t_3$, where the POIs with the same tag are colored with the same color. The pair values of each POI represent its keyword and rating.

{\bf POI Inverted Index (POI-II).} POI-II is designed to enable efficient retrieval of km-POIs for specific queries. Each keyword associated with a POI serves as a key in the index. Linked to each key is a list that includes all POIs tagged with that keyword. 

\begin{myDef}[Candidate POI set]
For a KATR query $q$ with a set of keywords $T_q=\lbrace t_1, t_2,…, t_m\rbrace$ and the entire set of POIs $\mathcal{P}$, a candidate POI set related to $q$, denoted as CP-Set, represents a collection of $m$ POIs from $\mathcal{P}$. Each POI in this set corresponds to a unique keyword from $T_q$, such that the keywords of the POIs in the CP-Set match $T_q$ exactly.
\end{myDef}

 \begin{myDef}[Route and Route Distance]
For any two vertices $v_s$ and $v_t$ in the graph $G$, a route $r$ between $v_s$ and $v_t$ is a sequence of vertices $\langle\texttt{v}_0=v_s$, $\cdots$, $\texttt{v}_l$, $\cdots$, $\texttt{v}_n=v_t\rangle$ such that $\forall l\in [1,n]$, $e_{i,j}\in E$ if $\texttt{v}_{l-1}=v_i$ and $\texttt{v}_l=v_j$. In this work, we only consider simple paths, i.e., paths with no repeat vertices. The distance of $r$ is defined as $Dis(r)$=$\sum\limits_{i=1}^n w_{i-1,i}$.
\end{myDef} 


{\bf Shortest distance.} For any two distinct vertices  $v_i$ and $v_j$ ($i\neq j$) in a given graph $G$, the shortest distance between $v_i$ and $v_j$, denoted by $SD(v_i, v_j)$, is the minimum value among all path distances connecting $v_i$ and  $v_j$ in $G$.

\begin{myDef}[candidate route]
For a given KATR query $q$ and a CP-Set, a candidate POI route for $q$ derived from the CP-Set, abbreviated as CP-Route, represents a route that commences at specified source and proceeds through the POIs within the CP-Set in any order. Its distance is defined as the sum of the shortest distances from the source to consecutive POIs on the given graph.
\end{myDef}




A CP-Route derived from a CP-Set not only includes POIs matching all keywords but also probably involves multiple km-POIs sharing the same keyword. Inspired by the work~\cite{zheng2015interactive}, our study incorporates only the highest rating POI for each keyword in the CP-Route, as defined below.

\begin{myDef}[candidate route score]\label{def:cr-score}
For the query $q (v_q,{T_q}=\lbrace t_1, t_2,…, t_m\rbrace)$ and a CP-Route $r_i$, let $\mathcal{P}_j$ represent the set of km-POIs matching the keyword $t_j\in{T_q}$ included in $r_i$. The score of $r_i$, $Score(r_i)$, is given by
$$Score(r_i)=-\alpha\times Dis(r_i)+\left(1-\alpha\right)\times\sum_{j=1}^{m}{\underset{v\in \mathcal{P}_j}{\max}{\, \lbrace v.\tau\rbrace}}$$ 
Here, $\underset{v\in \mathcal{P}_j}{\max}{\, \lbrace v.\tau\rbrace}$ is the highest rating of POI among $\mathcal{P}_j$ and $\alpha$ $(0\leq\alpha\leq 1)$ is a user-specified parameter.  
\end{myDef}

The above score is proportional to the cumulative highest ratings of POIs matching each query keyword along the route and inversely proportional to the route distance. The parameter $\alpha$ modulates the trade-off between rating and distance. A lower $\alpha$ favors high-rating POIs, allowing for the inclusion of distant popular POIs, while a higher $\alpha$ prioritizes proximity, favoring nearby POIs even if lower ratings. 
{POI ratings and road segment distances originate from different domains, potentially reducing the effectiveness of parameter $\alpha$. To address this, we normalize each POI rating and edge weight to standard ranges, setting the POI rating range ten times larger than that of edge weights, considering the typical tenfold ratio of edges to POIs in a route. This normalization prevents large disparities, keeping the values within the effective range of $\alpha$.}

\begin{myDef}[Keywords Aware Top-$k$ Route (KATR) Query]\label{def:katr-query}
On a graph $G({V}, E, W)$ with a set of POIs $\mathcal{P}$, the top-$k$ routes of {query $q (v_q,{T_q}=\lbrace t_1, t_2,…, t_m\rbrace)$} refers to a collection of $k$ CP-Routes, represented as $\mathcal{R}(q)=\lbrace r_1, r_2, \cdots, r_k\rbrace$, within graph $G$. These routes adhere to the condition $Score(r_i)\geq Score(r_{i+1})$ for $i\in[1,k-1]$, and for any route $r_j\notin\mathcal{R}(q)$, $Score(r_k)\geq Score(r_j)$.
\end{myDef}


 \begin{figure}[htpb!]
\centering
\vspace{-0.4cm}
\includegraphics[width=0.5\textwidth]{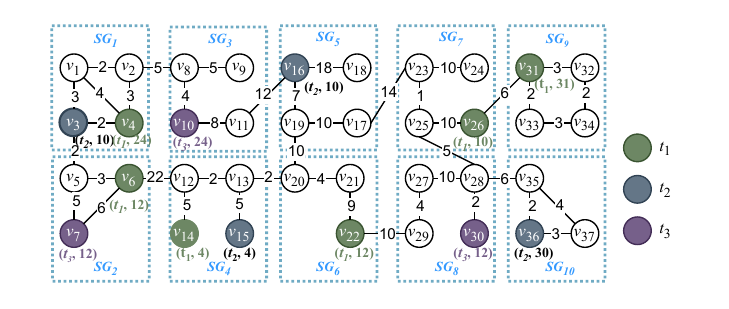}
\vspace{-0.6cm}
\caption{Subgraph Partitioning ($z$=4).}
\label{fig:SG}
 \vspace{-0.3cm}
\end{figure}

{The complexity of a KATR query is roughly determined by the number of potential route options. For a query $q$ with $m$ keywords, where each keyword $t_i$ corresponds to $|\mathcal{P}_i|$ POIs spread, these POIs can collectively form $\prod \limits_{i=0}^m{|\mathcal{P}_i|}$ CP-Sets related to $q$. Given that the POIs within each CP-Set can be visited in any sequence, each sequence represents a distinct CP-Route, making the number of CP-Routes for each CP-Set equal to $m!$. Consequently, the total number of CP-Routes can grow exponentially to $\prod \limits_{i=0}^m{|P_i|}\cdot m!$. This extensive candidate pool presents a significant challenge in efficiently identifying the top-$k$ routes.} 
















\begin{figure*}[]
    \centering
    \includegraphics[width=.95\linewidth]{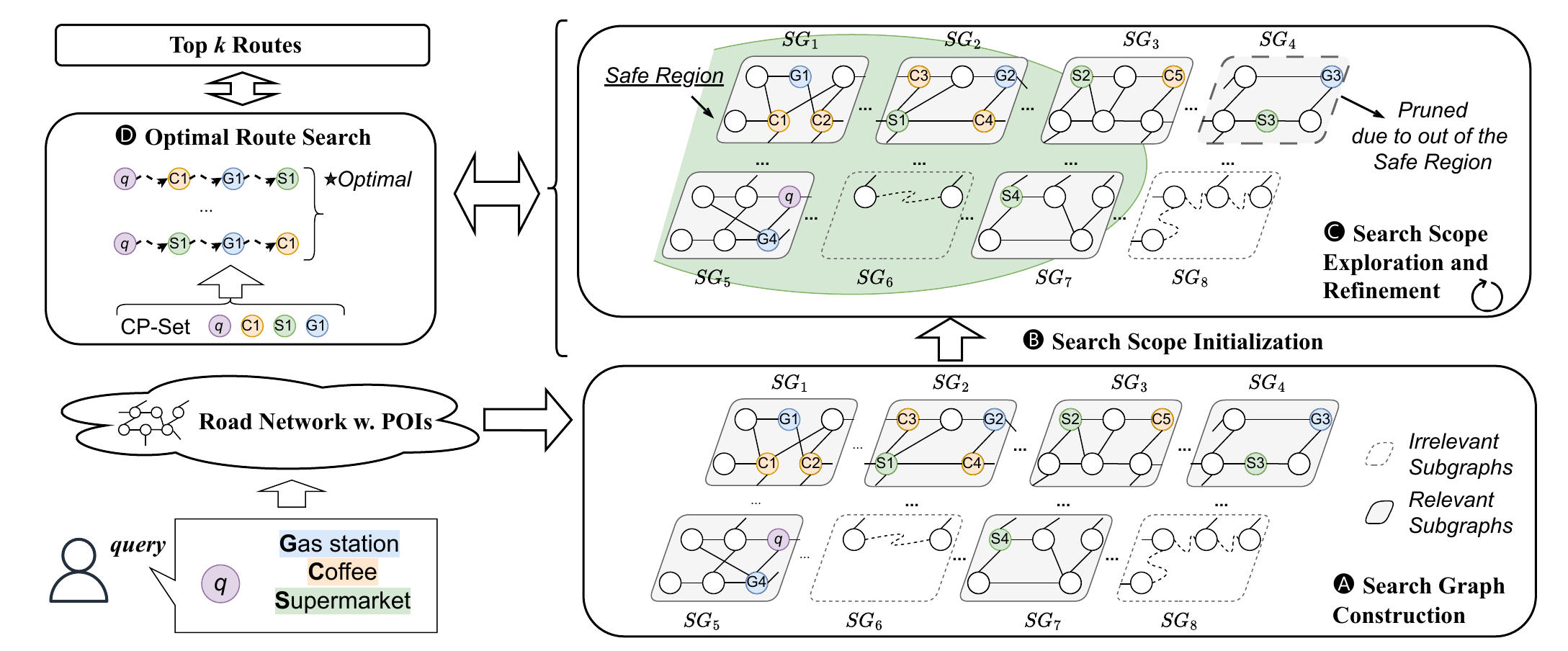}
    \vspace{-0.2cm}
    \caption{\small The KATR-Search Framework. For a given query $q$, we: \bcircled{A} Construct the Search Graph based on the road network to only include vertices that are needed by the query; \bcircled{B} Perform the first exploration step and find an upper bound for potential route distances; \bcircled{C} Explore within the search region and further refine the current search bound iteratively. CP-Sets are generated during exploration in both steps \bcircled{B} and \bcircled{C}, and we utilize \bcircled{D} to find the optimal route for a given CP-Set. 
    Edge weights are omitted for brevity, additionally, we use the first letter from each keyword to denote the POI that matches the corresponding keyword on the graph. 
    Better viewed in color. 
    }
    \label{fig:framework-overview}
    \vspace{-0.5cm}
\end{figure*}


\section{KATR-Search Framework}\label{sec:KATR-Search}
In this section, we introduce the Keyword Aware Top-$k$ Route Search framework (KATR-Search) to tackle KATR queries, aiming to efficiently identify $k$ optimal routes among a vast pool of potential routes by promptly narrowing the search space. 

\subsection{Preprocessing}
The KATR-Search partitions the graph $G$ into multiple subgraphs using graph partitioning algorithm like METIS~\cite{METIS}, utilizing these subgraphs as basic units to narrow the search space. Each subgraph contains no more than $N_s$ vertices, as defined in Definition~\ref{def:subgraph}.

\begin{myDef}[Subgraph]\label{def:subgraph}
A graph ${SG}$=(${V}'$, ${E}'$, ${W}'$) is a subgraph of the graph $G$= ($V, {E}, {W}$, ${P}$) iff (1) ${V}'\subseteq {V}$ and ${V}'\leq{N_s}$, (2) ${E}'\subseteq {E}$ $\land (e_{m, n}\in {{E}}'\to v_m, v_n\in {V}')$, and (3) ${W}'\subseteq {W}$ $\land (w_{m, n}\in {W}'\to e_{m, n}\in {E}')$.
\end{myDef}

 Within each subgraph, the shortest distance between any two vertices are precomputed and indexed.
The partitioned subgraphs are interconnected by \textit{external edges} that link \textit{border vertices}. A vertex is deemed a \textit{border vertex} if it has at least one adjacent vertex belonging to a different subgraph, while an edge connecting two border vertices from different subgraphs is classified as an \textit{external edge}. In KATR-Search, two border vertices connected by an external edge in two different subgraphs both record this external edge. Fig.~\ref{fig:SG} displays the partitioned subgraphs for the graph presented in Fig.~\ref{fig:original-Graph}, with a subgraph size threshold of 4. 

{\subsection{Overview of KATR-Search}\label{sec:KATR-Search:overview}
For a query $q(v_q,{T_q})$ with the query vertex $v_q$ and the set of query keywords $T_q$, KATR-Search aims to identify top-$k$ optimal routes passing through the POIs tagged with all keywords in $T_q$. 
As per Definitions \ref{def:cr-score} and \ref{def:katr-query}, routes in the top-$k$ results should either: (a) have a relatively low route distance, or (b) have a relatively high cumulative rating of POIs. Namely, candidate routes that have a high distance and low cumulative rating of POIs should be pruned early during exploration. Based on this observation, we employ an explore-and-bound paradigm. Specifically, we greedily explore the network while updating the bounds involving both the route distance and the cumulative ratings for effective early pruning and early stopping. Our framework consists of the following steps.

{\bf(A) Search Graph Construction:} Constructs Search Graph based on the given query, which consists of only vertices and POIs that could contribute to the top-$k$ routes utilizing graph partitioning. This step is a preliminary pruning step.
    
{\bf(B) Search Scope Initialization:} Performs coarse-grained pruning on the Search Graph utilizing the upper bound of the route distance, which consists of two steps: (1) explore: identify $k$ seed CP-Routes around $v_q$, and (2) bound: compute the upper bound (denoted as Safe Region) of the potential top-$k$ route distances based on the identified $k$ seed CP-Routes. All km-POIs that are out of the Safe Region will be pruned, significantly reducing the search space.
    
{\bf(C) Search Scope Exploration and Refinement:} Performs fine-grained pruning in two steps: (1) explore: explore CP-Routes within the Safe Region identified from step (B) to compute their scores, and (2) bound: based on the CP-Routes explored so far, compute score bounds for potential routes involving unprocessed POIs, and prune irrelevant POIs and CP-Sets involving these POIs.
    
{\bf(D) Optimal Route Search Module:} Searches for the optimal CP-Route for a given CP-Set among various candidates without exhaustively computing precise scores for all CP-Routes, which will be utilized during exploration from steps (B)~and~(C).
}

{Throughout the above steps, the Safe Region is continuously refined and dynamically shrinks as superior routes are revealed, while the ongoing refinement further enhances the efficiency of uncovering superior routes within the Safe Region. This iterative process persists until the Safe Region is fully explored, ultimately resulting in the identification of the final top-$k$ routes. Next, we discuss these steps in detail.}

{\subsection{Search Graph Construction}\label{subsec:query-graph}} 
{For a given query, except for the km-POIs corresponding to each keyword and necessary connections between these km-POIs, there are still extensive vertices irrelevant to the identification of top-$k$ optimal routes. Hence, KATR-Search introduces a \textbf{Search Graph} that eliminates irrelevant vertices and preserves crucial ones and their connections for each query, to delineate the search scope necessary for revealing the top-$k$ routes. The Search Graph customized for each query is built based on partitioned subgraphs as follows.}

{First, we identify km-POIs associated with each query keyword using the POI inverted index (POI-II). Since each POI has a subgraph identifier, we can quickly identify ``relevant subgraphs'' containing km-POIs, while those without are termed ``irrelevant subgraphs''. 

Next, we connect these subgraphs via their external edges between border vertices to form the Search Graph $G_s$. Specifically, each relevant subgraph is entirely included by $G_s$ while only border vertices of each irrelevant subgraph are included. Within each irrelevant subgraph, border vertices are directly connected with shortcuts. If an irrelevant subgraph contains the query vertex $v_q$, $G_s$ also includes the shortcuts between $v_q$ and each border vertex within that subgraph, ensuring the shortest path between vertices in $G_s$ remains consistent with that in the original graph $G$}.
These shortcuts are directly accessible, as the shortest distances between any two vertices within each subgraph are precomputed during the graph partitioning preprocessing. For example, in Fig. \ref{fig:framework-overview}, subgraph $SG_6$ is irrelevant, with no km-POI, while subgraph $SG_7$ is relevant. Instead of retaining all vertices in irrelevant subgraphs, only the border vertices and external edges are kept.

\subsection{Search Scope Initialization}\label{subsec:safe-region-build}
{With the Search Graph from Section~\ref{subsec:query-graph}, we next aim to find a scope that covers top-$k$ routes, enabling us to perform coarse-grained pruning based on the search scope, which involves two main steps. Initially, we wish to gradually explore the Search Graph, starting from $v_q$, to identify $k$ distinct candidate POI routes (CP-Routes) as seed routes. 
Subsequently, these $k$ seed CP-Routes are harnessed to delineate the Safe Region, which is typically much smaller than the Search Graph but encompasses the final top-$k$ routes.
}
\vspace{-0.1cm}
{\subsubsection{Identifying $k$ seed CP-Routes.} 
{
For any $k$ arbitrarily CP-Routes, where the lowest score among these routes is denoted as $SC_{min}$, the final top $k$ routes are guaranteed to have scores that are impossibly less than $SC_{min}$. In other words, the lowest score among any $k$ CP-Routes sets a baseline score for any other potential CP-Routes aspiring to be among the top-$k$.
Based on such observation, we propose to first identify such $k$ CP-Routes, denoted as seed $k$ CP-Routes, and then use them to set the search scope for the query.
}

{
Since any $k$ routes can serve this purpose, we employ a heuristic approach, involving two steps: (1) greedily identify $k$ distinct CP-Sets {starting with} $v_q$, and (2) select the optimal CP-Route from each CP-Set to form the $k$ seed CP-Routes. 
}

{\bf Identifying $k$ distinct CP-Sets.} 
We adopt a relaxation principle akin to Dijkstra's algorithm to gradually explore the Search Graph starting from the query vertex $v_q$. When encountering a km-POI, KATR-Search generates CP-Sets comprising this km-POI along with other previously identified km-POIs associated with different keywords. The exploration procedure, named DI-Expansion, halts once at least $k$ CP-Sets are obtained, as outlined in Algorithm~\ref{algorithm:DI-Exploration}. 

In DI-Exploration, a vertex $v_i$ is marked as ``processed'' concerning $v_q$ if $SD(v_i, v_q)$ (shortest distance between $v_i$ and $v_q$) on $G_s$ is established; otherwise, it remains ``unprocessed''. Initially, only $v_q$ is processed, leaving others unprocessed. For unprocessed vertices, the distances from those adjacent to $v_q$ are assigned their weights, whereas non-adjacent vertices receive infinity. A priority queue $Q_u$ only hold the unprocessed vertices with finite distances to $v_q$ and sort them in ascending order based on their distances to $v_q$. $Q_u$ is first filled with the adjacent vertices of $v_q$ (line 2-3) and subsequently, the algorithm iteratively dequeues the first vertex $v_f$, marks it as processed, and executes the subsequent steps.

{\bf $\langle {\bf 1}\rangle$} If $v_f$ is a km-POI matching the keyword $t_i$ ($i\in[1, m]$), processing involves two steps. First, $v_f$ is added to a set $\mathcal{P}_i$ containing km-POIs matching $t_i$. At this stage, If each keyword's km-POI set isn't empty, $v_f$ is combined with km-POIs in other sets to produce all new CP-Sets stemming from $v_f$. 
These newly generated CP-Sets are then appended to a set $\mathcal{I}$ (lines 5-10). Second, the Dijkstra's algorithm relaxation operation via  $v_f$ is conducted. Specifically, for each unprocessed adjacent vertex $v_a$ of $v_f$, $SD(v_q, v_j)$ is updated to $SD(v_q, v_f)+w_{f,a}$ if $SD(v_q, v_a)>SD(v_q, v_f)+w_{f,a}$, and $v_a$ is placed into $Q_u$ if it was not previously in $Q_u$ (lines 13-16). {\bf $\langle {\bf 2}\rangle$} If $v_f$ is not a km-POI, only the second step is executed. Repeat these steps until the Search Graph is fully traversed or the number of identified CP-Sets in $\mathcal{I}$ is no less than $k$ (lines 11-12).} 

\begin{algorithm}[!ht]
\caption{DI-Exploration}
\label{algorithm:DI-Exploration}
\SetKwData{Or}{\textbf{or}}
\DontPrintSemicolon
\KwIn {$G_s$, $v_q$, and query keywords $\lbrace t_1, \cdots, t_m\rbrace$}
\KwOut {Set $\mathcal{I}$ that contains at least $k$ distinct CP-Sets}
$\mathcal{P}_i$ contains km-POIs matching each keyword $t_i$($i\in[1,m]$)\;

  Initialize a priority queue $Q_u$\;
  
  Put all adjacent vertices of $v_q$ into $Q_u$\;

\While{$Q_u\neq \emptyset$}
{
$v_f$=$Q_u$.pop()\;

\If{$v_f$ is a km-POI matching $t_i$}
{
Insert $v_f$ into $\mathcal{P}_i$\;

\If{$\forall j \in [1, m] \setminus \{i\}$, $\mathcal{P}_j \neq \emptyset$}
{
Generate CP-Sets composed of $v_f$ and POIs from $\mathcal{P}_j$\; 

 Add newly generated CP-Sets into $\mathcal{I}$\;

 \If{$|\mathcal{I}|\geq k$}
{
Return $\mathcal{I}$\;
}
}
}
\For{$v_a$: unprocessed adjacent vertices of $v_f$}
{
\If{$SD(v_q,v_a)>SD(v_q, v_f)+w_{f,a}$}
{
$SD(v_q,v_a)$=$SD(v_q, v_f)+w_{f,a}$\;

Put $v_a$ into $Q_u$ if it was not in previously\;
}
}
}
\end{algorithm}

{\bf Selecting $k$ seed CP-Routes.} 
{Once $k$ CP-Sets are identified, we then select the optimal CP-Route from each CP-Set in $\mathcal{I}$ to form $k$ seed routes.} Since any two CP-Routes from the same CP-Set share the same POIs, the shortest one has the highest score. Therefore, the goal is to find the shortest CP-Route from each CP-Set. Considering computing the shortest distance for each CP-Route is computationally demanding, {an Optimal Route Search module is introduced to speedup this computation, which will be discussed in Section~\ref{subsec:EDRS}. We store the currently identified $k$ routes into $\mathcal{R}_c$, which will be updated progressively to include any new superior routes explored later until the top-$k$  routes are finalized}. 

\subsubsection{The Safe Region}\label{subsubsec:radius of safe-zone}

{After $k$ seed CP-Routes are identified, we aim to derive the search scope that is guaranteed to include top-$k$ CP-Routes.}
{Let $D_{ub}$ denote the upper bound on the route distance for all potential superior routes to the $k$ seed routes kept in $\mathcal{R}_c$. {The Safe Region, denoted as $SR$, can be viewed as a circular area centered around the query vertex $v_q$ with radius $D_{ub}$, where all vertices with shortest distances to $v_q$ not exceeding $D_{ub}$ will be included. 
We now discuss how to calculate this upper bound $D_{ub}$.
}

{{\bf Calculation of $D_{ub}$.} For any potentially superior CP-Route $r_x$ with the route distance $D_x$ and the cumulative POI rating $\tau_x$, its score $SC_x$ is given by 
\begin{equation}
    SC_x=-\alpha\cdot D_x+(1-\alpha)\cdot \tau_x
\end{equation}
based on Definition~\ref{def:cr-score}. Then we derive the route distance as:
\begin{equation}\label{eq:distance-upper-bound}
    D_x=\frac{(1-\alpha)\cdot \tau_x-SC_{x}}{\alpha}
\end{equation}
Clearly, the upper bound route distance $D_{ub}$ we aim to determine is the maximum conceivable value of $D_x$.  

Recall that the minimum score among the $k$ seed routes in $\mathcal{R}_c$, denoted by $SC_{min}$, set a performance baseline for all potential superior CP-Routes, signifying that the route $r_x$ with a score below $SC_{min}$ impossibly become a top-$k$ contender. Hence, we can infer that $SC_{x}\geq SC_{min}$. Further, we use $\tau_u$ to represent the maximum cumulative POI rating among all undiscovered CP-Sets that are readily identified based on POI-II, then $\tau_x\leq \tau_u$ must hold. {Substitute $SC_{x}$ with $SC_{min}$, and $\tau_x$ with $\tau_u$, in Equation~\ref{eq:distance-upper-bound}, we have}

\begin{equation}
\vspace{-0.1cm}
    D_x\leq\frac{(1-\alpha)\cdot \tau_u-SC_{min}}{\alpha}
\end{equation}
Hence, a tight upper bound on the route distance for all potentially superior routes can be computed as
\begin{equation}\label{equ:upper-bound-route-distance}
D_{ub}=\frac{(1-\alpha)\times \tau_u- SC_{min}}{\alpha}
\end{equation}
This formula conclusively establishes the boundary of the Safe Region, ensuring comprehensive coverage of all routes that could potentially achieve top-$k$ status.


\begin{theorem}\label{theo:safe-region}
    For a given query $q(v_q,T_q)$ on the Search Graph $G_s$, the identified Safe Region $SR$ must encompass the top-$k$ optimal routes across the entire Search Graph w.r.t. $q$.
\end{theorem}
This theorem can be easily proved by contradiction, so the detailed proof is omitted here.

{{\bf Safe Region establishment.} After identifying the upper bound route distance $D_{ub}$, we initialize the Safe Region. Since the Safe Region encompasses the final top-$k$ optimal routes, we just need to consider the unknown CP-Sets within the Safe Region. And because some km-POIs with high ratings on the Search Graph may no longer be included within the Safe Region, the parameter $\tau_u$ in Equation~\ref{equ:upper-bound-route-distance}, which indicates the maximum cumulative rating among the yet-to-be-discovered CP-Sets within the Safe Region, is likely to diminish.  This reduction in $\tau_u$ helps in further narrowing the Safe Region. 

However, the exact value of $\tau_u$ within the Safe Region is unknown due to our lack of knowledge regarding all km-POIs present in this area, as much of the Safe Region remains uncharted. To address this, we conduct an efficient range exploration on the Search Graph by traversing only the border vertices of each subgraph using Dijkstra's algorithm, {starting from} $v_q$ with {an upper bound on the route distance} of $D_{ub}$. This method provides us with the shortest distances from $v_q$ to the border vertices of subgraphs within $D_{ub}$. By identifying subgraphs that intersect with the Safe Region—those containing at least one border vertex within $D_{ub}$—we can then recalculate $\tau_u$ by considering only potential CP-Sets composed of km-POIs within these intersected subgraphs. This approach helps refine our estimate of $\tau_u$ and ensures we consider all km-POIs within the Safe Region.

If $\tau_u$ remains unchanged, $D_{ub}$ stays the same, thereby establishing the Safe Region. If $\tau_u$ decreases, $D_{ub}$ is further reduced, tightening the Safe Region and potentially lowering $\tau_u$ again. Note that in each iteration of shrinking the Safe Region, there is no need to conduct another range search to identify intersected subgraphs with the Safe Region, as the border vertices within the smaller $\tau_u$ are already obtained. This process continues until the maximum potential cumulative rating $\tau_u$ among CP-Sets within the Safe Region no longer decreases with the reduction in $D_{ub}$.

\begin{figure}[htbp!]
  \centering
\vspace{-0.3cm}
\begin{subfigure}{1.1\linewidth}
      \centering   
\includegraphics[width=\textwidth]{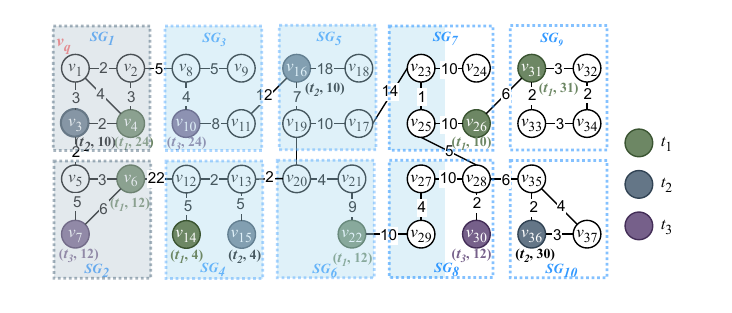}
    \vspace{-0.8cm}
    \caption{First Round}
    \label{subfig:initial-safe-region}
    \end{subfigure}
    \vspace{-0.2cm}
\begin{subfigure}{1.1\linewidth}
      \centering   
\includegraphics[width=\textwidth]{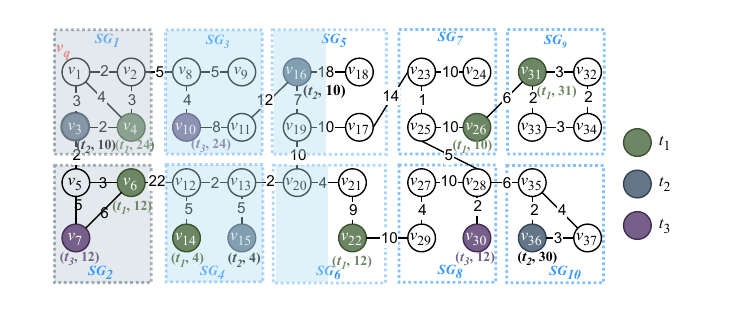}
    \vspace{-0.8cm}
    \caption{Second Round}\label{subfig:safe-region-second-round}
    \label{}
    \end{subfigure}  
\vspace{-0.2cm}
\caption{Process of Establishing the Safe Region}
 \vspace{-0.2cm}
\end{figure}

\begin{example}\label{example:safe-region-built}
    Fig.~\ref{subfig:initial-safe-region} shows an example to establish the Safe Region for the given query $q$ with $v_q$ being $v_1$ and keywords are $\lbrace t_1, t_2, t_3\rbrace$. Initially, we conduct the DI-Exploration strategy from $v_1$, sequentially processing $v_2$, $v_3$, $v_4$, $v_5$, $v_8$, $v_6$, $v_7$. When $v_7$ is processed, we get $\mathcal{P}_1=\lbrace v_4,v_6\rbrace$, $\mathcal{P}_2=\lbrace v_3\rbrace$, and $\mathcal{P}_3=\lbrace v_7\rbrace$. Then two CP-Sets $\lbrace v_4, v_3, v_7\rbrace$ and $\lbrace v_6, v_3, v_7\rbrace$ can be produced, and their optimal routes are $r_1=\lbrace v_1, v_4, v_3, v_5, v_7\rbrace$  with distance $D_{r_1}=13$ and rating $\tau_{r_1}=46$, and $r_2=\lbrace v_1, v_3, v_5, v_6, v_7\rbrace$ with distance $D_{r_2}=14$ and rating $\tau_{r_2}=34$ respectively. Assuming $\alpha =0.5$, we infer that $SC(r_1)=16.5$ and $SC(r_2)=10$ based on the score equation from Definition~\ref{def:cr-score}. Since the highest cumulative rating among unknown CP-Sets on the entire Search Graph is 85, derived from $(v_{31}\in \mathcal{P}_1, v_{36}\in\mathcal{P}_2$, $v_{10}\in\mathcal{P}_3)$, the radius of the Safe Region, $D_{ub}$, can be deduced as 65 based on Equation~\ref{equ:upper-bound-route-distance}.  The Safe Region is visualized by the shaded area, with intersected subgraphs $SG_1, SG_2, \cdots, SG_8$.

    Since $SG_9$ and $SG_{10}$ are excluded from the Safe Region, highest accumulative rating of POIs among unknown CP-Sets in the Safe Region becomes 58, derived from ($v_4\in\mathcal{P}_1$, $v_{3}\in\mathcal{P}_{2}$,   $v_{10}\in\mathcal{P}_3$), as $v_{31}\in\mathcal{P}_1$ and $v_{36}\in\mathcal{P}_2$ are not included by Safe Region. At this point,  $D_{ub}$ is updated as 52, contracting the Safe Region as shown in Fig.~\ref{subfig:safe-region-second-round}. With no further change in the highest a accumulative rating among unknown CP-Sets between the current and the last versions of the Safe Region, the Safe Region is established. 
 \end{example}}

\vspace{-0.2cm}
\subsection{Search Scope Exploration and Refinement}\label{subsubsec:refine-safe-zone}
Once the Safe Region is identified, we progress to uncover potentially superior routes within this area by advancing DI-Exploration along the frontier of the already explored region. As better routes are revealed, the value of $SC_{min}$ gradually increases, which in turn decreases $D_{ub}$, the radius of the Safe Region, according to $D_{ub}=\frac{(1-\alpha)\cdot \tau_u-SC_{min}}{\alpha}$ (Equation~\ref{equ:upper-bound-route-distance}). This leads to the continuous refinement of the Safe Region. Next, we detail the process of unveiling superior routes, driving this refinement process further.


{\subsubsection{Batch Pruning of km-POIs}\label{subsec:bp-subgraph}
Limiting our search to the Safe Region significantly reduces computational resources. However, evaluating each km-POI within this region for its potential to contribute to superior routes remains resource-intensive. Given that the Search Graph consists of multiple subgraphs and that km-POIs in many subgraphs may not be part of the top-$k$ routes, a subgraph-based pruning strategy emerges. This strategy treats each subgraph as a unit and estimates a score upper bound for all potential CP-Routes, which may involve other subgraphs, passing through at least one POI within the subgraph. This score is termed the ``score upper bound for the subgraph''. If this score upper bound does not exceed $SC_{min}$, the score of the current $k^{th}$ best route, {all km-POIs within the subgraph can be safely eliminated.} 


{For a subgraph $SG_i$, we now discuss how the score upper bound, denoted as  $SC_{SG_i}$, can be computed. For any potential CP-Route $P_y$ involves at least one POI within $SG_i$, we have
}

\vspace{-0.3cm}
\begin{equation}\label{eq:upper-bound-score-subgraph}
\vspace{-0.1cm}
    SC_y=-\alpha\cdot D_y+(1-\alpha)\cdot \tau_y
\end{equation}

{
The score upper bound of $SG_i$ is then the upper bound value of $SC_y$. To estimate the upper bound value of $SC_y$, we need to determine the minimum possible value of distance, $D_y$ (the first term in the Equation~\ref{eq:upper-bound-score-subgraph}), and the maximum possible value of POI rating, $\tau_y$ (the second term in the Equation~\ref{eq:upper-bound-score-subgraph}).
}

{\em Bound the minimum value of $D_y$.} We use $D_{lb}$ to denote the minimum value of $D_y$, that is, the route distance lower bound for any CP-Route including unprocessed km-POIs within $SG_i$. If $SG_i$ represents a partially unexplored subgraph (shortened as pu-subgraph), with certain vertices already processed, $D_{lb}$ is set as the minimum among the shortest distances from $v_q$ to all processed vertices within $SG_i$. Conversely, for a fully unexplored subgraph (tu-subgraph), where the closest border vertex of $SG_i$ to $v_q$ is the next vertex for processing, $D_{lb}$ is determined by the shortest distance from $v_q$ to the nearest border vertex of $SG_i$. Regardless of being a pu-subgraph or tu-subgraph, it is clear that $D_{lb}$ cannot exceed the shortest possible route distance involving POIs in $SG_i$. 

{\em Bound the maximum value of $\tau_y$.} Let $\tau_{max}$ denote the maximum value of $\tau_y$, the highest cumulative POI rating among the CP-Sets containing at least one unprocessed km-POI in $SG_i$. To do this, we select the km-POI with the highest rating for each keyword to form a CP-Set $CPS_{hr}$. If $CPS_{hr}$ includes at least one unprocessed km-POI in $SG_i$, then the cumulative rating of POIs in $CPS_{hr}$ is $\tau_{sub}$. Otherwise, we create a new CP-Set  $CPS'_{hr}$ by replacing a km-POI $v'_u$ in $CPS_{hr}$ with an unprocessed km-POI $v_u$ in $SG_i$, such that 1) $v_u$ and $v'_u$ correspond to the same keyword; 2) the cumulative rating of POIs in $CPS'_{hr}$ is maximized. At this moment, the cumulative rating of POIs in $CPS'_{ph}$ is $\tau_{max}$.

{Substitude $D_y$ with $D_{lb}$, and $\tau_{y}$ with $\tau_{max}$ in Equation~\ref{eq:upper-bound-score-subgraph}, we have }
\vspace{-0.3cm}
\begin{equation}
\vspace{-0.05cm}
    SC_y\leq=-\alpha\cdot D_{lb}+(1-\alpha)\cdot \tau_{max}
\end{equation}
Hence, the upper score bound for $SG_i$, $SC_{SG_i}$, is determined as $SC_{SG_i} = -\alpha \times D_{lb} + (1-\alpha) \times \tau_{max}$. 

If $SC_{SG_i} \leq SC_{min}$, it signifies that potential routes incorporating km-POIs within $SG_i$ are unlikely to rank among the top-$k$ routes. In this case, we bypass the subgraph by only visiting its border vertices, avoiding further exploration within $SG_i$. 
Otherwise, the km-POIs in $SG_i$ cannot be dismissed outright, and we proceed to evaluate CP-Sets derived from these km-POIs. 
} 

{\begin{example}
Continuing from Example~\ref{example:safe-region-built}, DI-Exploration processes to visit the border vertex $v_{12}$ of subgraph $SG_4$ after exploring $SG_1$, $SG_2$, and $SG_3$ within the Safe Region. It evaluates the score upper bound for $SG_4$. Since $SG_4$ is entirely unexplored, the lower bound route distance for potential routes involving km-POIs in $SG_4$ is set to the shortest distance from $v_q$ to the nearest border vertex $v_{12}$ of $SG_4$, which is $SD(v_q, v_{12})=30$. Moreover, the maximum cumulative POI rating among all CP-Sets is 85 deriving from the CP-Set $\lbrace v_{31}(t_1), v_{36}(t_2), v_{10}(t_3)\rbrace$, not including any unprocessed km-POI in $SG_4$. Hence, by substituting $v_{36}$ with $v_{15}$ to form the CP-Set $\lbrace v_{31}(t_1), v_{15}(t_2), v_{10}(t_3)\rbrace$, which has the highest cumulative POI rating of 59 among all CP-Sets involving km-POIs in $SG_4$. Assuming $\alpha=0.5$, the score upper bound for $SG_4$, denoted as $SC_{SG_4}$, is computed as $-0.5 \times 30 + 0.5 \times 59 = 9.5$. As $SC_{SG_4} < SC_{r_2}=10$ (discussed in Example~\ref{example:safe-region-built}), it implies that the km-POI within $SG_4$ can be safely ignored. Hence, DI-Exploration bypasses $SG_4$ by only visiting its border vertices from $v_{12}$ to $v_{13}$.
\end{example}}
\vspace{-0.3cm}

\begin{figure}[htp!]
\centering
\includegraphics[width=0.5\textwidth]{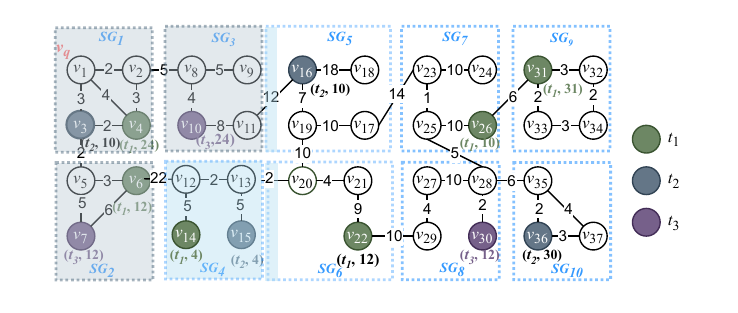}
 \vspace{-0.2cm}
\caption{Safe Region Refinement}\label{SG}
\vspace{-0.3cm}
\end{figure}

\vspace{-0.2cm}
{\subsubsection{Unpromising CP-Sets elimination}\label{subsubsec:CPSet-prune}
When a subgraph $SG_i$ cannot be bypassed, it's necessary to evaluate the CP-Sets formed by the km-POIs within it. Many of these CP-Sets don't have potential to contribute to the top-$k$ CP-Routes. Therefore, it's crucial to efficiently filter them out without fully calculating their potential scores. To tackle this, KATR-Search quickly estimates an upper bound score for all potential CP-Routes in a given CP-Set. If this upper bound does not exceed $SC_{min}$, the CP-Set can be safely discarded.

We now discuss estimating the score upper bound for a CP-Set $CPS_i$. All potential CP-Routes within $CPS_i$ share the same set of POIs, leading to an identical cumulative rating of POIs denoted as $\tau_r$. Let $D_{r_i}$ denote the distance of any potential CP-Route $r_i$ on the Search Graph, then the score of $r_i$ is $SC_{r_i}=-\alpha\cdot D_{r_i}+(1-\alpha)\cdot \tau_r$. Since the exact computation of $SC_{r_i}$ is infeasible due to the unknown $D_{r_i}$, we derive an upper bound on $SC_{r_i}$ by {substituting} $D_{r_i}$ with $ED_{r_i}$, the ``Euclidean distance'' of CP-Route $r_i$. This {substitution} is valid because the Euclidean distance of any CP-Route cannot exceed its distance on the Search Graph. The Euclidean distance of a CP-Route refers to the sum of the straight-line distances from $v_q$ to consecutive POIs along this route, efficiently computed based on their coordinates.

Next, we compute the Euclidean distance for every CP-Route within the CP-Set $CPS_i$ and determine the minimum Euclidean distance ($ED_m$) among them. Since $ED_m$ cannot exceed the actual distance of any CP-Route among $CPS_i$, the score upper bound for $CPS_i$, $SC_{CPS_i}$, is set as $SC_{CPS_i}=-\alpha\cdot ED_m+(1-\alpha)\cdot \tau_r$, signifying no potential CP-Route among this CP-Set can achieve higher score than $SC_{CPS_i}$. Consequently, if $SC_{CPS_i}\leq SC_{min}$ holds, then no CP-Route in $CPS_i$ can surpass the current top-$k$ routes, enabling the removal of $CPS_i$. Otherwise, further assessment is necessary to identify the optimal CP-Route within $CPS_i$ with {the Optimal Route Search} module (Section~\ref{subsec:EDRS}). If the optimal CP-Route surpasses $SC_{min}$, KATR-Search will use this route to replace the least favorable route in $\mathcal{R}_c$.} 

{\subsubsection{Safe Region refinement}\label{subsubsec:safe-region-refine}
In the above process, once a superior route is discovered to substitute the least favorable route in $\mathcal{R}_c$, the Safe Region is refined accordingly. In particular, $SC_{min}$ is updated to reflect this new superior route's inclusion. Then the Safe Region $D_{ub}$ is adjusted based on the revised $SC_{min}$ as per Equation~\ref{equ:upper-bound-route-distance}, such that $D_{ub}=\frac{(1-\alpha)\times \tau_u- SC_{min}}{\alpha}$. As $SC_{min}$ increases with the addition of superior routes, $D_{ub}$ will correspondingly decrease, narrowing the Safe Region. 
This refinement process iterates continuously as more superior routes are discovered, progressively narrowing the Safe Region until it is fully examined.}
{\subsection{Optimal Route Search: The EDRS Algorithm}\label{subsec:EDRS}
Given a CP-Set comprising $m$ POIs, there are $m!$ distinct sequences in which these POIs can be visited, each forming a specific CP-Route. {To reduce the computational cost and perform effective route search, we introduce the Euclidean Distance Referenced Search (EDRS) algorithm to promptly pinpoint the optimal one, which forms the Optimal Route Search module in our framework.}





{\bf{The Main Idea} of EDRS.} EDRS leverages the Euclidean distances tied to CP-Routes as a reference to exclude less relevant routes by adhering to the following guidelines: (1) For any given CP-Route, its distance on the Search Graph (referred to as ``graph distance''), is generally proportional to its ``Euclidean distance''
. Thus, CP-Routes with smaller Euclidean distances {should be} prioritized, as they indicate potentially shorter graph distances. (2) {Since the graph distance of any CP-Route will never be less than its Euclidean distance, a CP-Route can be immediately eliminated if its Euclidean distance exceeds the minimum graph distance among evaluated CP-Routes, as it cannot yield a shorter route.}



{\bf Procedure of EDRS.} {For most CP-Sets, the Euclidean distances of CP-Routes are already calculated during the elimination of CP-Sets (Section~\ref{subsubsec:CPSet-prune}). For the remaining CP-Sets, EDRS computes the Euclidean distances for their CP-Routes. Then, for each CP-set, it sorts their CP-Routes based on their Euclidean distances in ascending order, and computes their graph distances sequentially to identify the shortest one. Specifically, it tracks the currently identified smallest graph distance, denoted as $GD_s$. For each examined CP-Route $r_i$, if its Euclidean distance $ED_i$ is not smaller than $GD_s$, signifying that its graph distance $GD_i$ cannot be less than $GD_s$. This rationale holds true for all CP-routes not yet evaluated, as their Euclidean distances cannot be smaller than $ED_i$. Hence, the assessment of the remaining POI sequences can be halted, establishing that the CP-Route with $GD_s$ is the shortest one within this CP-Set. If $ED_i<GD_s$, EDRS computes the graph distance $GD_i$ of $r_i$. If $GD_s>GD_i$, EDRS updates $GD_s$ as $GD_i$ and records the route $r_i$. The process repeats until encountering the terminal condition or exhausting all sequences.}


{\bf Graph distance computation.} For a given CP-Route, the query vertex $v_q$ and the POIs on the route are treated as ``pivot vertices''. Calculating the graph distance involves finding the shortest distance between each pair of adjacent pivot vertices. If two adjacent pivot vertices are within the same subgraph, their indexed shortest distance is readily available. If they are in different subgraphs, algorithms like A* are used to find the shortest distance through the Search Graph. Since shortcuts from each pivot vertex to the border vertices of its subgraph are indexed, the computation focuses only on the pivot and border vertices of the traversed subgraphs, ensuring efficiency.}

{\bf Top-$k$ optimal routes identification.} EDRS  serves as a cornerstone in searching for the optimal route from each assessed CP-Set as outlined in Sections~\ref{subsec:safe-region-build} and \ref{subsubsec:refine-safe-zone}. If the identified optimal route surpasses the current $k^{th}$ best route, it replaces the least favorable route in $\mathcal{R}_c$, which holds the current top-$k$ routes. Hence, the $k$ routes in $\mathcal{R}_c$ constitute the final results once the Safe Region is fully explored.

\subsection{Discussions}\label{subsec:discussion}

{\bf Time cost.} The time complexity of KATR-Search for processing a KATR query with $m$ keywords is $O((N_v+N_e) \log N_v + \omega \cdot N_v\times \beta \cdot {N_P}^{(m-1)} \times \gamma \cdot m!)$, consisting of two main components. 

{\em Safe Region exploration:}  KATR-Search navigates through the Search Graph within the Safe Region with a complexity of $O((N_v + N_e) \log N_v)$, where $N_v$ and $N_e$ are the number of vertices and edges in the Safe Region respectively.

{\em Candidates evaluation:} Suppose there are $\omega$ percent of $N_v$ km-POIs in the Safe Region to be evaluated after the subgraph-based pruning. For each km-POI, KATR-Search generates ${N_P}^{(m-1)}$ CP-Sets to be evaluated, with $N_p$ representing the average number of km-POIs per keyword. After unpromising CP-Set pruning, assume $\beta$ percent of CP-Sets are retained for further evaluation. For each CP-Set, $\gamma$ percent of CP-Routes ($m!$) require graph distance computation. Thus, the evaluation cost is $(\omega \cdot N_v \times \beta \cdot {N_P}^{(m-1)} \times \gamma \cdot m!)$, where graph distance computation for each route is considered a unit overhead. 




{{\bf Maximum search scope.} Let's estimate the maximum search scope of KATR-Search, focusing on the approximate area of the Safe Region. For simplicity, we assume the graph distance closely matches the Euclidean distance between vertices.

Assume each query keyword brings $N_P$ km-POIs uniformly distributed across the Search Graph of size $z$. POI ratings range from $\tau_l$ to $\tau_h$, and both ratings and edge weights are normalized to the range $(0, 1]$. For a query centered on vertex $v_q$ with $m$ keywords, there are $N_P^m$ CP-Sets distributed over the Search Graph. Thus, the area containing $k$ distinct CP-Sets around $v_q$, called the ``seed area'', is approximately $\frac{k}{N_P^m} \cdot z$ and is assumed to be circular. The maximum distance among $k$ seed routes from the CP-Sets is considered the diameter of this seed area, calculated as $2\sqrt{\frac{k}{{N_P}^m \cdot \pi} \cdot z}$. The combined POI ratings along these routes must be at least $m \cdot \tau_l$, so the minimum score for these routes is $SC_{min} = -\alpha \times 2\sqrt{\frac{k}{N_P^m \cdot \pi} \cdot z} + (1-\alpha) \cdot m \cdot \tau_l$.

With the maximum total rating in a CP-Set being $m \cdot \tau_h$, the Safe Region radius is approximately $D_{ub}=\frac{(1-\alpha)\times m\cdot{(\tau_h-\tau_l)}+2\sqrt{\frac{k}{N_P^m \cdot \pi} \cdot z}}{\alpha}$. Assuming the Safe Region is circular, its size is roughly $z_s\approx \pi \cdot \left(\frac{(1-\alpha)\times m\cdot{(\tau_h-\tau_l)}+2\sqrt{\frac{k}{N_P^m \cdot \pi} \cdot z}}{\alpha}\right)^2$.

We verify this analysis with experimental results. The pruning ratio by Safe Region is $\omega=\frac{z_s}{z}$. In one experiment with $\pi=3.14$, $m=4$, $k=4$, $N_P=10$, $\alpha=0.5$, and $\tau_h-\tau_l=0.5$, we calculate $\omega=\frac{3.14 \cdot (2.5+2\sqrt{\frac{z}{7850}})^2}{z}$. For $z=1,000,000$ with both dimensions of the Search Graph area at 1000, $\omega$ is approximately 0.2\%. This estimate matches our experimental findings in Fig.~\ref{expfig:KTAR-search-prune}, showing that CP-Sets within Safe Region account for only 0.45\% of all CP-Sets in NY.}

\section{Experiment}\label{sec:experiment}


We implemented all algorithms in Java and conducted the experiments on a PC with a 4.4 GHz core and 16 GB of RAM.

{\bf Datasets.} We used four real-world datasets: Oldenburg (OL), California (CA), Beijing (BJ), and New York City (NY), summarized as Table~\ref{Statistics of Datasets}. Each dataset represents a road network with POIs distributed on it. The CA dataset's road network and POIs are sourced from \cite{turchetto2023random}, while the road networks of OL,  BJ, NY, and FLA are sourced from \cite{turchetto2023random}, \cite{Range2023WSDM}, and \cite{li2023finding} respectively. The corresponding POI sets were obtained from OpenStreetMap \cite{vargas2020openstreetmap} and mapped to the nearest vertex in their respective road networks. 

\begin{table}[htbp]
\vspace{-0.3cm}
\tiny
\centering
\caption{\small{Statistics of Datasets}}\label{Statistics of Datasets}
\vspace{-0.1in}
\resizebox{\linewidth}{!}{
\begin{tabular}{lllll}
\hline
\makecell[cl]{Road Networks}&  \makecell[cl]{\# Vertices} &  \makecell[cl]{\# Edges}  & \makecell[cl]{\# Keyword tags}&
\makecell[cl]{\# POIs}\\
\hline
OL & 6,105 & 7,035 & 26 & 2,404 \\
CA & 21,048 & 41,693 & 64 & 15,635\\
BJ & 115,436 & 215,344 & 23 & 25,075 \\
 NY & 264,346 & 733,846 & 135 & 30,382 \\
 FLA & 1,070,376 & 2,712,798 & 384 & 73,472\\
 \hline
\end{tabular}}
\vspace{-0.15in}
\end{table}


{\bf Baselines.} We incorporate DAPrune~\cite{li2023finding}, ROSE-GM~\cite{zhu2022top}, OSScaling~\cite{cao2012keyword}, and StarKOSR~\cite{liu2018finding} as baselines. The details of each approach have been discussed in Section~\ref{sec:relatd work}. Considering KATR-Search accommodate the query without a specific destination, we created adapted versions of OSScaling and ROSE-GM, called ND-OSScaling and ND-ROSE-GM, to support route planning in this context. The characteristics of these benchmarks are summarized in Table~\ref{tab:baseline}.  


\begin{table}[h]
\newcommand{\tabincell}[2]{\begin{tabular}{@{}#1@{}}#2\end{tabular}}
\centering
\vspace{-0.2cm}
\caption{\small{Approaches in Evaluation}}\label{tab:baseline}
\vspace{-0.3cm}
\smallskip\noindent
\resizebox{1\linewidth}{!}{
\begin{tabular}{|c|c|c|c|c|}
\hline
{\bf Algorithm}  & {\bf  Flexible Visiting Order} & {\bf Diverse POI Rating} & {\bf Flexible Distance Budget} & {\bf Flexible Destination}\\
\hline
{ DAPrune} & { $\checkmark$} & { $\times$} & { $\checkmark$}& { $\times$}\\
\hline
{ ROSE-GM} & { $\times$}  & { $\checkmark$} & { $\checkmark$}& { $\times$}\\
\hline
{ ND-ROSE-GM} & { $\times$}  & { $\checkmark$} & { $\checkmark$} & { $\checkmark$}\\
\hline
{ StarKOSR} & { $\times$} & { $\times$} & { $\checkmark$}  & { $\times$}\\
\hline
{ OSSCaling} & { $\checkmark$} & { $\checkmark$} & { $\times$} & { $\times$}\\
\hline
{ND-OSSCaling} & {$\checkmark$}  & { $\checkmark$} & { $\times$}& { $\checkmark$}\\
\hline
{ Ours} & { $\checkmark$}  & { $\checkmark$} & { $\checkmark$} & { $\checkmark$}\\
\hline
\end{tabular}
}
\vspace{-0.3cm}
\end{table}
{Among these methods, ROSE-GM and StarKOSR establish a route distance budget to restrict the search space. In our comparative evaluation, the distance budget is set to be five times the maximum distance of the top-$k$ routes identified by our proposal. This ensures a large enough space for these baselines to identify the optimal routes based on their scoring functions.}

\begin{table*}[h!]
\vspace{-0.3cm}
\centering
\caption{Evaluation on Pruning Efficiency of KATR-Search on NY}\label{tab:prune-efficiency}
\vspace{-0.1in}
\begin{tiny}
\resizebox{0.9\linewidth}{!}{
\begin{tabular}{llllllllll}
\hline
\makecell[cl]{}&  \makecell[cl]{{$N_{SG}$({\em RN})}} &  \makecell[cl]{{$N_{SG}$({\em SR})}}  & \makecell[cl]{{$N_{SG}$({\em BP})}}&
\makecell[cl]{{$N_{CPS}$({\em RN})}}&
\makecell[cl]{{ $N_{CPS}$({\em SR})} }&
\makecell[cl]{{$N_{CPS}$({\em BP})} }&
\makecell[cl]{{$N_{CPR}$({\em SR})}}&
\makecell[cl]{{$N_{CPR}$({\em EDBS})}}\\
\hline
$\mu$=3 & 854 & 81 & 10 & 43200 & 48 & 21 & 443 & 156\\
$\mu$=4 & 943 & 97 & 13 & 261360 & 120 & 54 & 2630 & 917\\
$\mu$=5 & 1067 & 111 & 17 & 2874960 & 430 & 278 & 20960 & 7336\\
\hline
$k$=2 & 943 & 73 & 6 & 261360 & 48 & 19 & 1472 & 472\\
$k$=4 & 943 & 97 & 13 & 261360 & 120 & 54 & 2630 & 915\\
$k$=6 & 943 & 118 & 19 & 261360 & 216 & 105 & 3813 & 1342\\
\hline
$\alpha$=0.2 & 943 & 109 & 15 & 261360 & 178 & 93 & 35505 & 13740\\
$\alpha$=0.4 & 943 & 104 & 13 & 261360 & 145 & 70 & 28404 & 10566\\
$\alpha$=0.6 & 943 & 97 & 13 & 261360 & 120 & 54 & 21040 & 7406\\
\hline
\end{tabular}}
\end{tiny}
\vspace{-0.1in}
\end{table*}

{\subsection{Integrating LLM with KATR-Search}
To verify feasibility of integrating KATR-Search with LLM, we have implemented an agent that encapsulates KATR-Search as an external component of LLM, utilizing function calling techniques.
This provides an interface for users to input query keywords or conversations and receive optimal routes across the relevant POIs. Figure \ref{fig:llm-agent} illustrates the framework of this agent. 

In this agent, we define two functions {\bf POI-tags} and {\bf KATR-Search} as tools. The detailed codes of tools description is attached in the Appendix. When the agent receives a user submitted KATR query, it sends the query labeled ``KATR'' to LLM (step $\textcircled{1}$). Once LLM recognizes a KATR query, it first instructs the agent to invoke the function POI-tags to get the entire set of POI tags from our dataset (steps $\textcircled{2}$ and $\textcircled{3}$). It then matches the query keywords with the POI tags and identifies the set of POI tags involved in the query as the parameter for the KATR-Search algorithm. Subsequently, LLM prompts the agent to invoke the function {KATR-Search} that identifies and returns the top-$k$ optimal routes to LLM (steps $\textcircled{4}$ and $\textcircled{5}$). Finally, the LLM presents the optimal routes to the user in an easily understandable format (step $\textcircled{6}$). Fig.~\ref{} shows the effect of this agent.}

\begin{figure}[htpb!]
\centering
\includegraphics[width=0.5\textwidth]{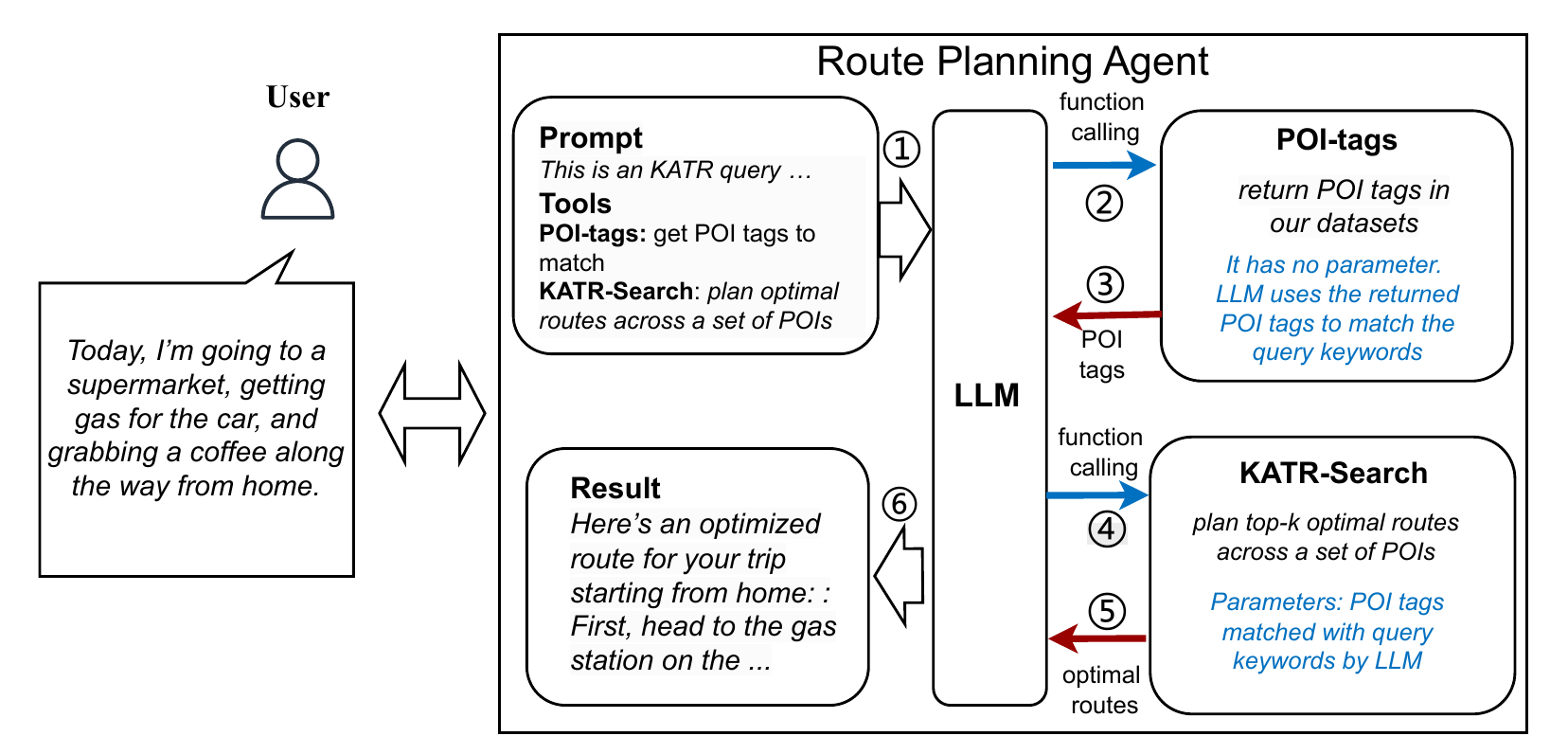}
\vspace{-0.6cm}
\caption{Route Planning Agent Framework.}
\label{fig:llm-agent}
\vspace{-0.4cm}
\end{figure}

{To evaluate whether LLMs can efficiently handle KATR queries without integrating our proposed KATR search algorithm, we compare the performance of two LLMs, ChatGPT-o3-mini-high and Deepseek, and our constructed agent in processing such queries. Since LLMs struggle to process KATR queries over large road networks, we extract a connected sub-network from New York City consisting of approximately 1,000 vertices and POIs to serve as our testing dataset. In this evaluation, we run the same set of queries, with $k=2$, on both LLMs and our agent.

 Fig.~\ref{LLM-comparison-query-time} shows that our agent employing the KATR-Search algorithm processes queries much faster than the two LLMs we tested. This is because the LLMs rely on low efficient heuristic search principle to enumerate the potential optimal routes, which introduces substantial overhead. Fig.~\ref{LLM-comparison-route-score} demonstrates that 
 the routes found by LLM+KATR-Search consistently achieve higher scores than those determined by the two LLMs alone. This discrepancy arises because
the LLMs cannot exhaustively enumerate all potential routes to identify the exact optimal ones and instead only generate approximate routes.}

\vspace{-0.2cm}
\begin{figure}[htb!]
  \centering
  \captionsetup{font={scriptsize}}  
\begin{subfigure}{0.48\linewidth}
      \centering   
\includegraphics[width=\textwidth]{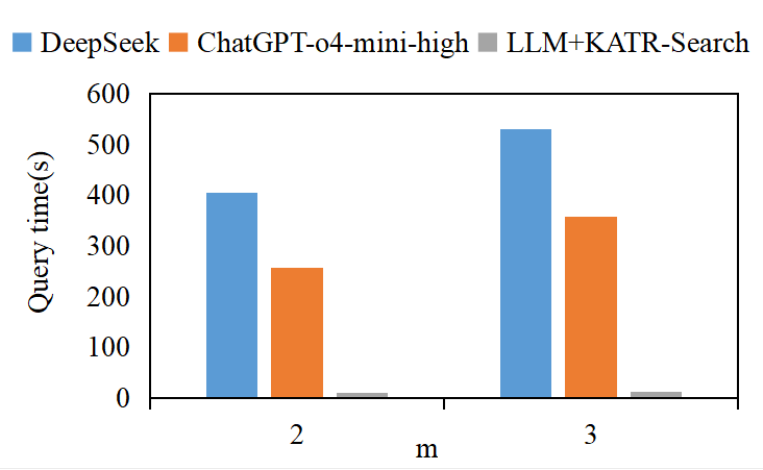}
    \vspace{-0.6cm}
    \caption{Query Time Comparison}\label{LLM-comparison-query-time}
    \end{subfigure}      
\begin{subfigure}{0.48\linewidth}
      \centering   
\includegraphics[width=\textwidth]{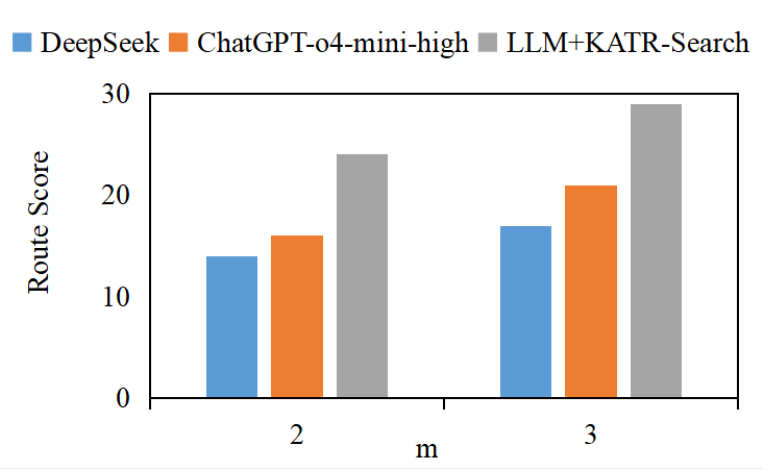}
    \vspace{-0.6cm}
    \caption{Route Score Comparison}\label{LLM-comparison-route-score}
    \end{subfigure}  
\vspace{-0.2cm}
\caption{Comparison with LLMs}
 \vspace{-0.4cm}
\end{figure}

\subsection{KATR-Search Evaluation}
In this section, we first evaluate the pruning effectiveness of KATR-Search by varying the parameters $k$ (number of optional routes), $m$ (number of query keywords), and $\alpha$ (the weight of rating and distance in the score function). We then test its query performance and conduct an ablation study to assess the impact of different pruning components in our methodology.
\subsubsection{Pruning Efficiency}
{To visually illustrate  the pruning effectiveness of various pruning strategies including the safe region, the km-POI batch pruning strategy, and the EDRS algorithm, Table~\ref{tab:prune-efficiency} displays  the values for $N_{SG}$, $N_{CPS}$, and $N_{CPR}$, which represent the  actual counts of relevant subgraphs (referred to as {\em SG}), candidate POI sets ({\em CP-Set}), and candidate POI routes ({\em CP-Route}) at various stages of exploration. This includes the initial state on the entire road network, the pruned state using the safe region, the further pruned state via the km-POI batch pruning strategy, and the final state after applying the EDRS algorithm. These status are denoted by $RN$, $SR$, $BP$, and $EDRS$, respectively. In Table~\ref{tab:prune-efficiency}, the first column shows the parameter settings. The next three columns present the number of relevant subgraphs ($N_{SG}$) in the unpruned road network ($RN$), within the initialized safe region ($SR$), and remaining after the km-POI batch pruning ($BP$), respectively. For instance, when $\mu=5$, there are 1,067 relevant subgraphs in total; only 111 (about 10\%) are within the safe region, and just 17 (about 1\%) require further exploration after km-POI batch pruning, highlighting the high effectiveness of these pruning methods. Similarly, the remaining columns emphasize the significant performance of these pruning strategies on CP-Sets and CP-Routes.}




To further demonstrate the pruning efficiency of KATR-Search, we present the remaining percentages of subgraphs, CP-Sets, and CP-Routes at various pruning stages for each dataset, as shown in Fig.~\ref{subfig:Pru-subraph} to \ref{subfig:Pru-CPRoutes}. Fig.~\ref{subfig:Pru-subraph} illustrates that, for each dataset, the percentage of subgraphs with km-POIs within the initialized Safe Region is below 15\% of the total subgraphs with km-POIs. Fig.~\ref{subfig:Pru-CPSets} shows that CP-Sets within the initialized Safe Region constitute less than 1.5\% of the total CP-Sets. After subgraph-based pruning and unpromising CP-Sets elimination, less than 1\% of CP-Sets undergo actual processing, indicating the pruning effectiveness of KATR-Search on CP-Sets. Fig.~\ref{subfig:Pru-CPRoutes} indicates that the EDRS algorithm prunes nearly 70\% of CP-Routes within the Safe Region for each dataset, undergoing the efficiency of this pruning strategy.

\begin{figure}[htp!]
  \centering
\captionsetup{font={scriptsize}}
\begin{subfigure}{0.48\linewidth}
      \centering   
\includegraphics[width=\textwidth]{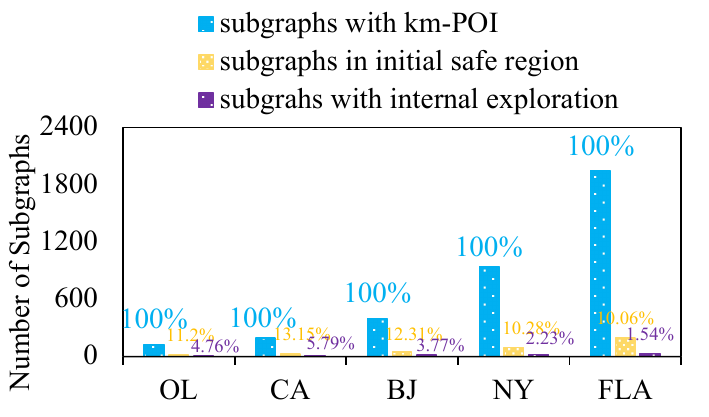}
    \vspace{-0.6cm}
    \caption{Subgraphs Pruning }\label{subfig:Pru-subraph}
    \end{subfigure}      
\begin{subfigure}{0.48\linewidth}
    \centering   
\includegraphics[width=\textwidth]{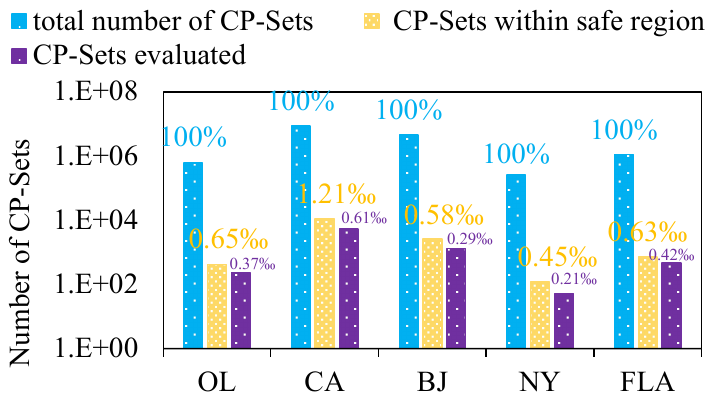}
    \vspace{-0.6cm}
    \caption{CP-Sets Pruning }\label{subfig:Pru-CPSets}
    \end{subfigure}
\begin{subfigure}{0.48\linewidth}
      \centering   
\includegraphics[width=\textwidth]{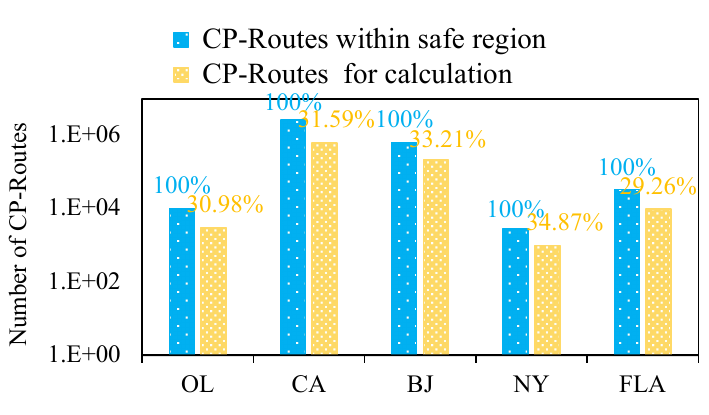}
    \vspace{-0.6cm}
    \caption{CP-Routes Pruning}\label{subfig:Pru-CPRoutes}
    \end{subfigure}  
\begin{subfigure}{0.48\linewidth}
    \centering   
\includegraphics[width=\textwidth]{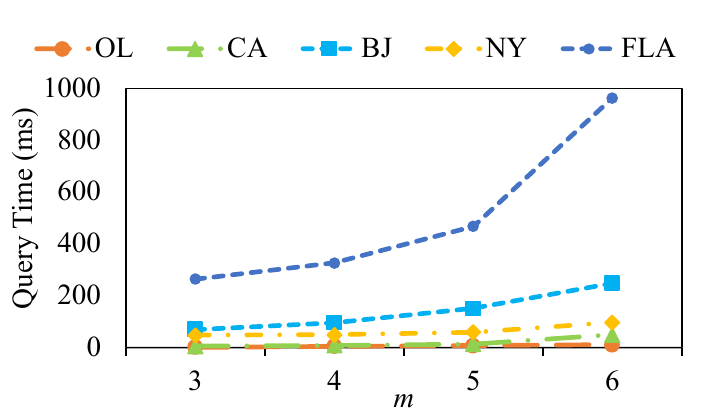}
    \vspace{-0.6cm}
    \caption{Query Time w.r.t. $m$}\label{Response-Time-mu}
    \end{subfigure}
 \begin{subfigure}{0.48\linewidth}
      \centering   
\includegraphics[width=\textwidth]{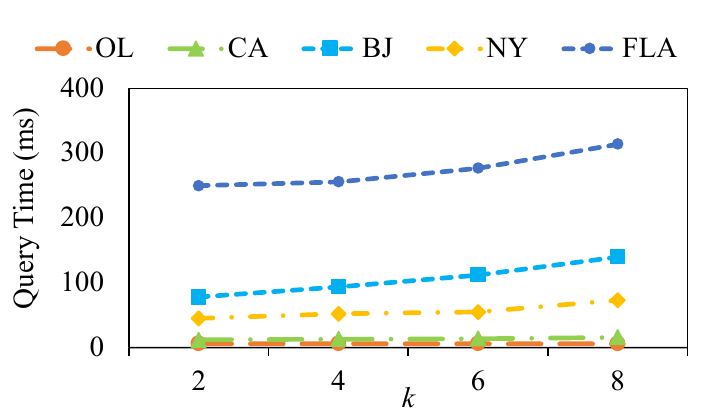}
    \vspace{-0.6cm}
    \caption{Query Time w.r.t. $k$}\label{Response-Time-k}
    \end{subfigure}   
\begin{subfigure}{0.48\linewidth}
      \centering   
\includegraphics[width=\textwidth]{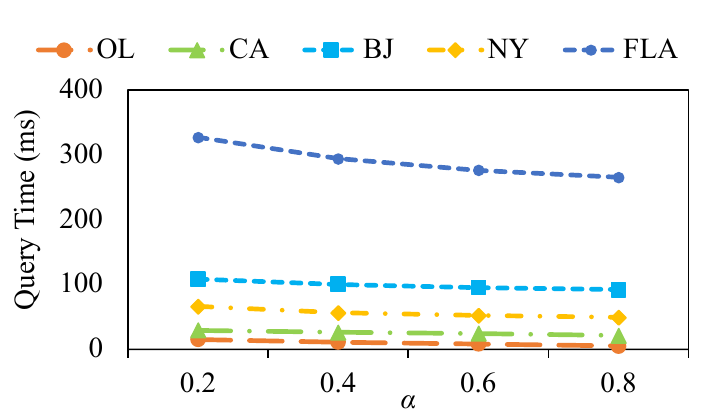}
    \vspace{-0.6cm}
    \caption{Query Time w.r.t. $\alpha$}\label{Response-Time-alpah}
    \end{subfigure} 
\caption{Query Performance of KATR-Search}\label{expfig:KTAR-search-prune}
 \vspace{-0.4cm}
\end{figure}

\subsubsection{Query efficiency}
We first analyze the query time of KATR-Search with varying parameters $k$, $m$, and $\alpha$ across different datasets. In this test, we process 1,000 queries with query vertices distributed across the entire road network and report the average query time. The results are shown in Fig.~\ref{Response-Time-mu}-\ref{Response-Time-alpah}. 
As depicted in Fig.~\ref{Response-Time-mu}, when $k$ and $\alpha$ are held constant, the query time increases almost linearly with $m$, the number of query keywords. Despite an exponential increase in candidate sets with $m$, the linear growth in query time underscores the efficiency of KATR-Search's pruning strategies. Similarly, Fig.~\ref{Response-Time-k} shows that query times also increase linearly as $k$ rises, but at a much slower rate compared to the growing number of optimal routes to be identified, underscoring KATR-Search's scalability with respect to $k$. Fig.~\ref{Response-Time-alpah} shows that query times decrease linearly as $\alpha$ increases, as a larger $\alpha$ emphasizes route distance in scoring, thus enhancing pruning efficiency. This is supported by the data in the last four rows of Table~\ref{tab:prune-efficiency}, which shows higher pruning efficiency reduces as $\alpha$ increases. 

Next, we perform an ablation experiment to assess the effectiveness of each pruning component in KATR-Search. We do this by creating variants of KATR-Search without Safe Region-based pruning (KATR-NoSR), subgraph-based pruning (KATR-NoSG), and EDRS-based pruning (KATR-NoED). We compare the query times of KATR-Search and these variants on NY, as shown in Fig.~\ref{Ablation-mu}-\ref{Ablation-mu-alpha}. KATR-Search consistently demonstrates superior performance, with its advantage becoming more pronounced as $m$ and $k$ values increase, as shown in Fig.~\ref{Ablation-mu} and \ref{Ablation-k}. This is because larger $m$ and $k$ generate more candidate results, amplifying the advantage of a comprehensive pruning strategy. In Fig.~\ref{Ablation-mu-alpha}, both KATR-Search and its variants show reduced query times as $\alpha$ increases. The performance gap between KATR-Search and its variants narrows slightly because larger $\alpha$ values reduce the candidate options, diminishing the relative advantage of KATR-Search's superior pruning. {In Fig.~\ref{subfig:ablation-POI-number}, we compare the performance with varying numbers of POIs from the NY dataset. We observe that the query time for KATR-Search and other algorithms increases with more POIs, but KATR-Search exhibits a slower rate of increase due to its pruning capability.} 

\begin{figure}[htb!]
  \centering
  \captionsetup{font={scriptsize}}  
\begin{subfigure}{0.48\linewidth}
      \centering   
\includegraphics[width=\textwidth]{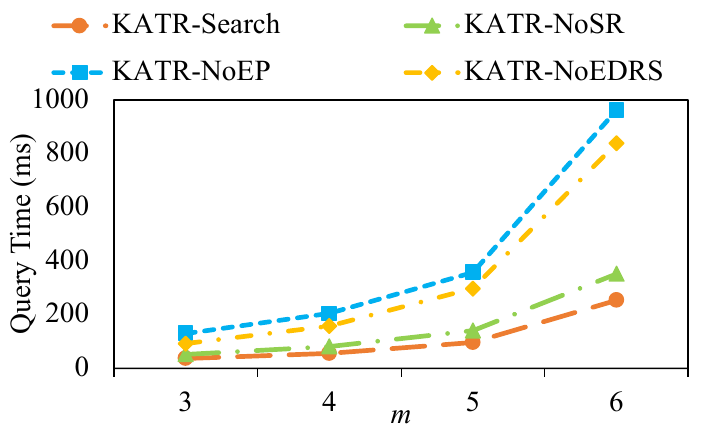}
    \vspace{-0.6cm}
    \caption{Ablation Evaluation w.r.t. $m$}\label{Ablation-mu}
    \end{subfigure}      
\begin{subfigure}{0.48\linewidth}
      \centering   
\includegraphics[width=\textwidth]{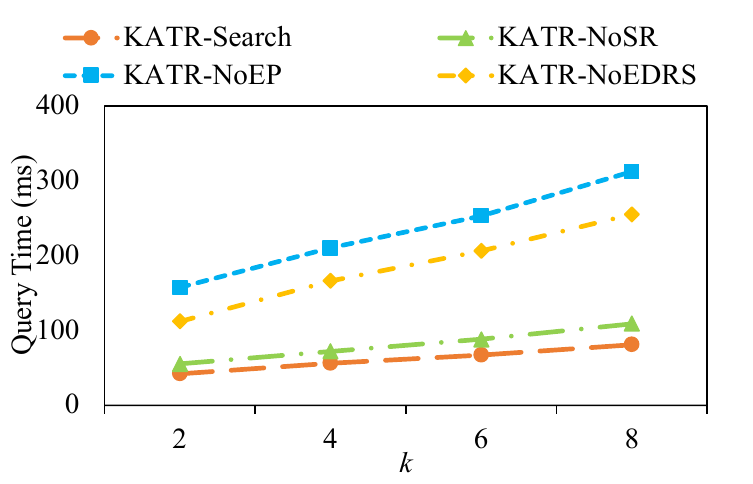}
    \vspace{-0.6cm}
    \caption{Ablation Evaluation w.r.t. $k$}\label{Ablation-k}
    \end{subfigure}  
\begin{subfigure}{0.48\linewidth}
      \centering   
\includegraphics[width=\textwidth]{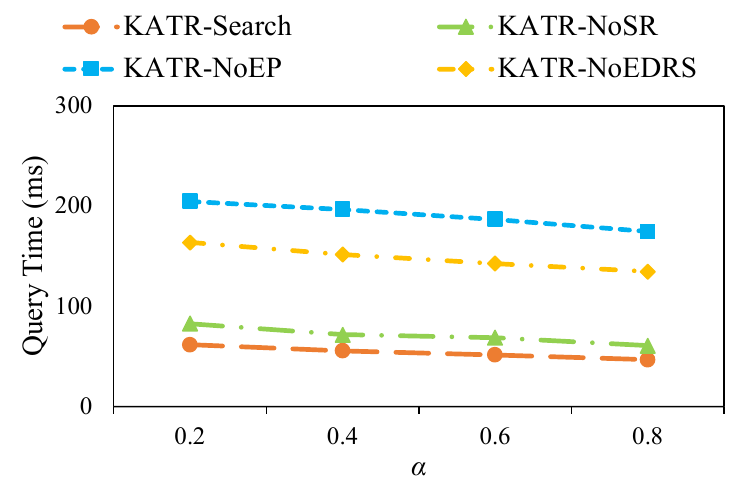}
    \vspace{-0.6cm}
    \caption{Ablation Evaluation w.r.t. $\alpha$}\label{Ablation-mu-alpha}
    \end{subfigure}
    \begin{subfigure}{0.48\linewidth}
      \centering       
      \includegraphics[width=\textwidth]{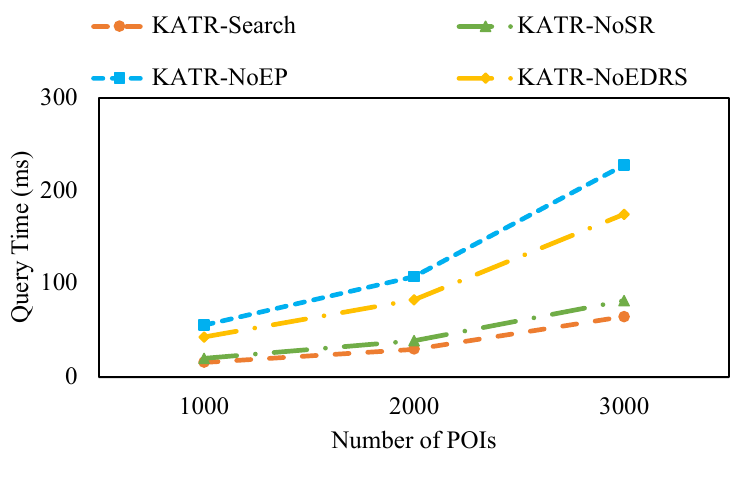}
    \vspace{-0.6cm}
    \caption{Ablation Evaluation w.r.t. \#POIs}\label{subfig:ablation-POI-number}
    \end{subfigure}
\vspace{-0.2cm}
\caption{Ablation Evaluation of KATR-Search}
 \vspace{-0.2cm}
\end{figure}

\begin{figure*}[htb!]
  \centering
  \captionsetup{font={scriptsize}}
\begin{subfigure}{0.2\linewidth}
      \centering   
\includegraphics[width=\textwidth]{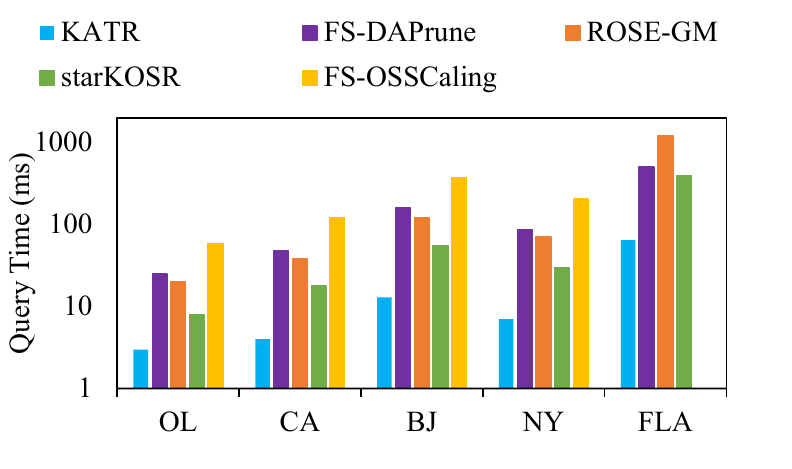}
      \vspace{-0.6cm}
      \caption{Fixed-order KATR w.r.t. datasets}\label{expfig:comparison-all-datasets}
      \end{subfigure}    
 \begin{subfigure}{0.2\linewidth}
      \centering   
\includegraphics[width=\textwidth]{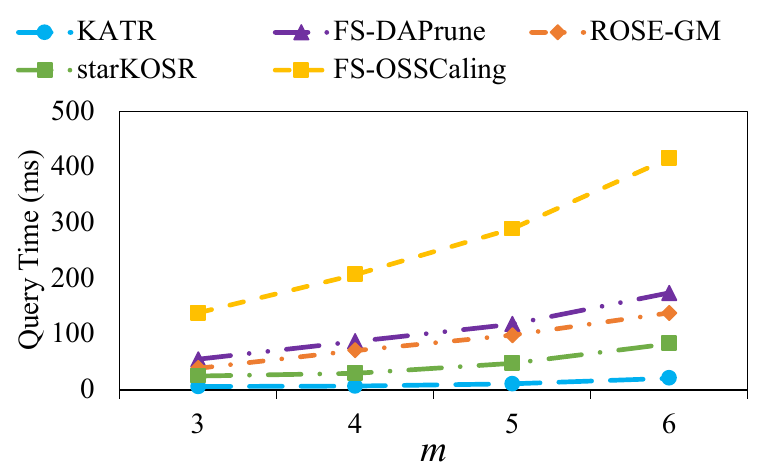}
      \vspace{-0.6cm}
      \caption{Fixed-order KATR w.r.t. $m$}\label{expfig:comparision-mu}
      \end{subfigure}  
\begin{subfigure}{0.19\linewidth}
      \centering   
\includegraphics[width=\textwidth]{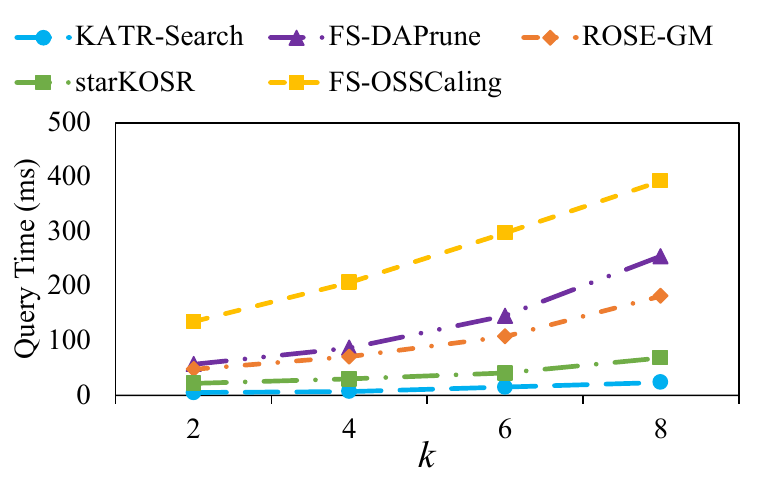}
      \vspace{-0.6cm}
      \caption{Fixed-order KATR w.r.t. $k$}\label{expfig:comparision-k}
      \end{subfigure}   
 \begin{subfigure}{0.19\linewidth}
      \centering   
      \includegraphics[width=\textwidth]{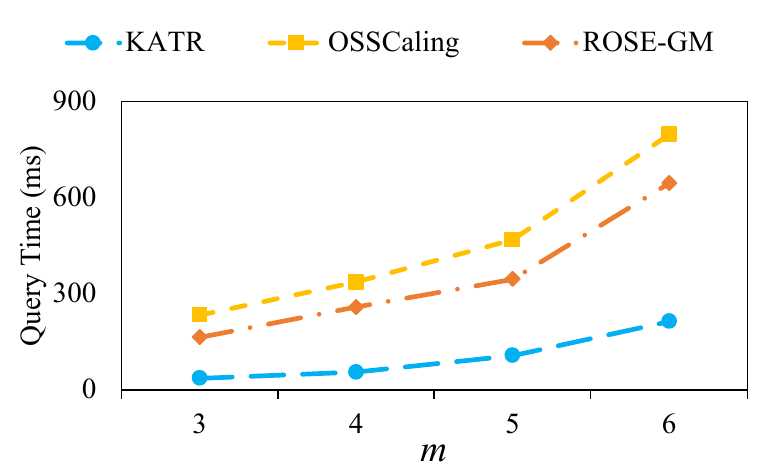}
      \vspace{-0.6cm}
      \caption{Fixed-order KATR (Diverse Rating) w.r.t. $m$}\label{expfig:comparion-unsequenced-mu}
      \end{subfigure}  
\begin{subfigure}{0.20\linewidth}
      \centering   
      \includegraphics[width=\textwidth]{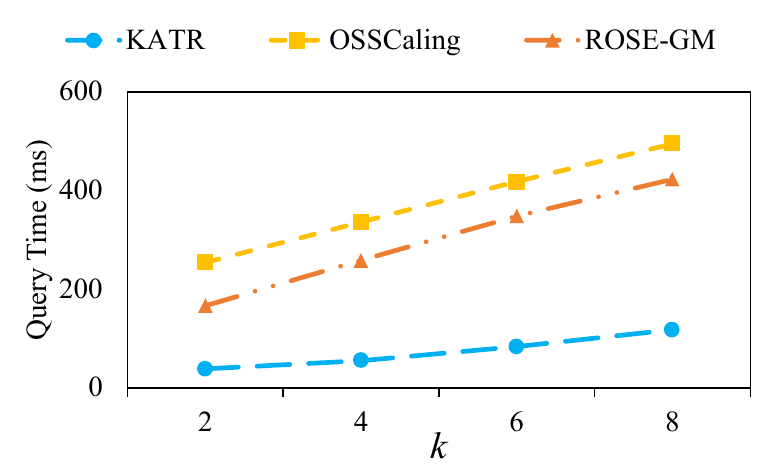}
      \vspace{-0.6cm}
      \caption{Fixed-order KATR (Diverse Ratings) w.r.t. $k$}\label{expfig:comparion-unsequenced-k}
      \end{subfigure}  
 \begin{subfigure}{0.2\linewidth}
      \centering   
      \includegraphics[width=\textwidth]{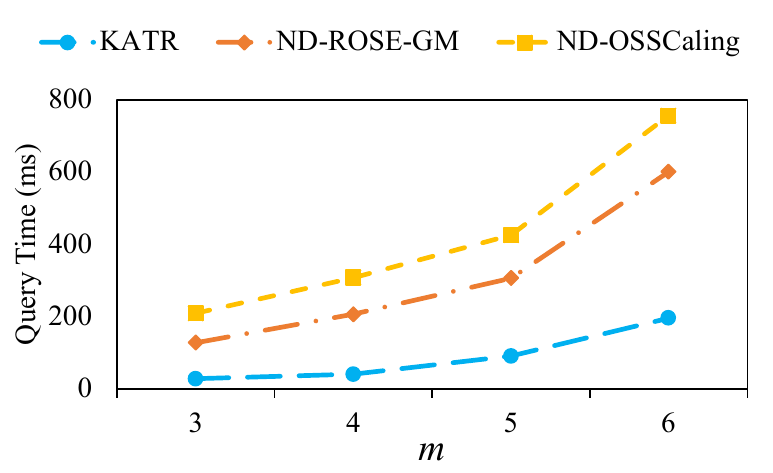}
      \vspace{-0.6cm}
      \caption{Fixed-order KATR (No Destination) w.r.t. $m$}\label{expfig:unsequenced-no-destination-mu}
      \end{subfigure}  
\begin{subfigure}{0.2\linewidth}
      \centering   
      \includegraphics[width=\textwidth]{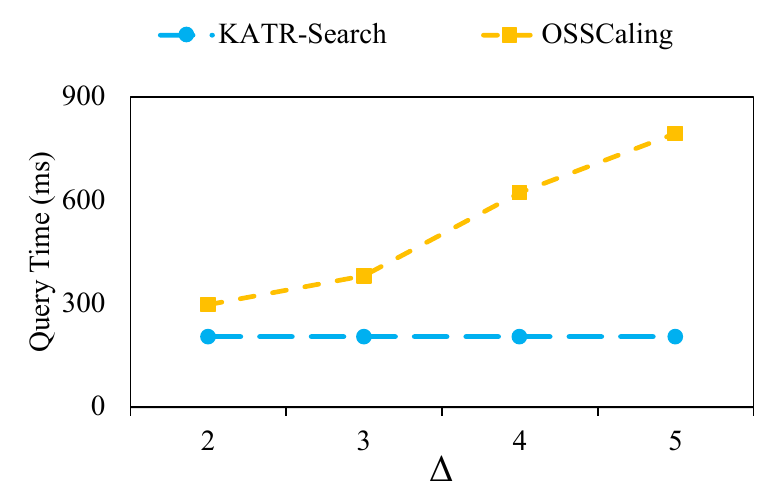}
      \vspace{-0.6cm}
      \caption{Distance budget $\Delta$ varying}\label{expfig:distance-budget-delta}
      \end{subfigure} 
    \begin{subfigure}{0.19\linewidth}
      \centering   
      \includegraphics[width=\textwidth]{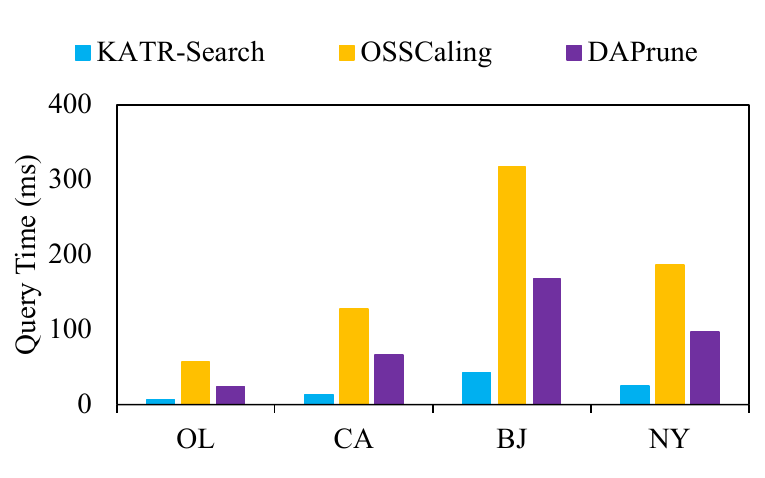}
      \vspace{-0.6cm}
      \caption{Flexible-order KATR w.r.t. Datasets}\label{expfig:global-datasets}
      \end{subfigure}    
 \begin{subfigure}{0.19\linewidth}
      \centering   
      \includegraphics[width=\textwidth]{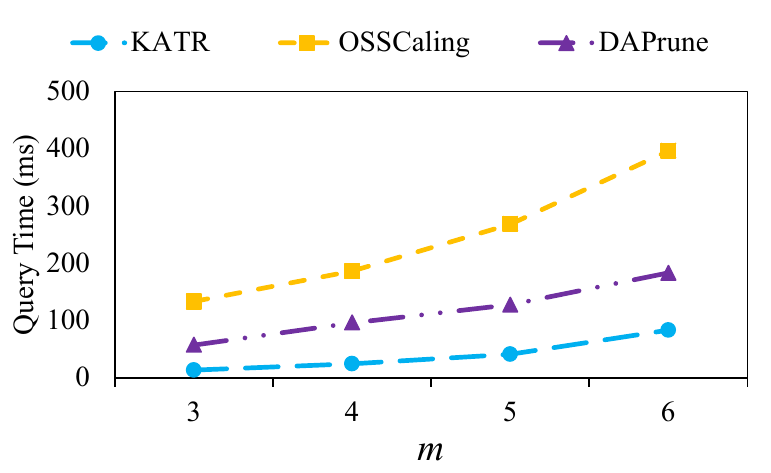}
      \vspace{-0.6cm}
      \caption{Flexible-order KATR w.r.t. $m$}\label{expfig:global-mu}
      \end{subfigure}  
\begin{subfigure}{0.20\linewidth}
      \centering   
      \includegraphics[width=\textwidth]{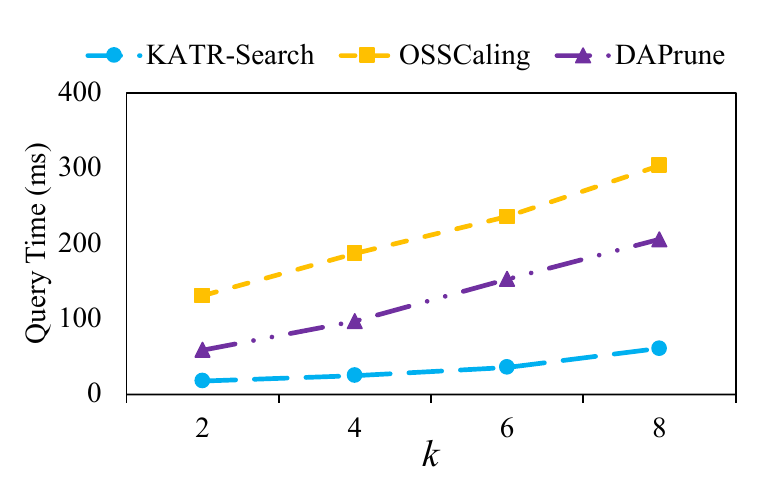}
      \vspace{-0.6cm}
      \caption{Flexible-order KATR w.r.t. $k$}\label{expfig:global-k}
      \end{subfigure} 
\vspace{-0.2cm}
\caption{
\label{exp:query-cost}
{Comparison with Baselines}
}
 \vspace{-0.5cm}
\end{figure*}

\vspace{-0.2cm}
\subsection{Comparison with Baselines}
In this section, we compare the query performance of KATR-Search with the baselines DAPrune, OSSCaling, ROSE-GM, StarKOSR, and their respective adaptations. 

\subsubsection{Comparison in fixed visiting order}
ROSE-GM and StarKOSR are tailored for fixed POI visiting order whereas DAPrune and OSScaling are not. For fairness, we configure DAPrune and OSSCaling to identify optimal routes following the keyword sequence, referring to the modified versions as ``FS-DAPrune'' and ``FS-OSSCaling'' respectively. 

{\bf Identifical POIs' ratings.} In this evaluation, we assign the same rating to all POIs and introduce an unique destination in the queries. We evaluate query times of all approaches on each dataset using default parameter values. The results, presented in Fig.~\ref{expfig:comparison-all-datasets}, show that KATR-Search consistently outperforms all other baselines. Except for StarKOSR, which enforces a strict rating threshold to reduce the number of involved km-POIs, our approach outperforms other baselines by one to two orders of magnitude. In this test, OSSCaling, which relies on indexed shortest distances between any two vertices, requires nearly 1 terabyte of memory on FLA—exceeding our device's capacity—and is therefore not shown for FLA.

We also compare all methods' query performance varying values of $m$ and $k$. Fig.~\ref{expfig:comparision-mu} and \ref{expfig:comparision-k} indicate that KATR-Search excels over all baselines, with its advantage increasing as $m$ and $k$ grow. This advantage stems from the exponential rise in partial candidate routes for baselines using the PNE principle as $m$ and $k$ increase, impacting baseline performance.

{\bf Varying POIs' ratings.} Now we compare KATR-Search with baselines under varied POI ratings. Since only ROSE-GM and OSSCaling consider the diversity in ratings of POIs, we focus our comparison on these two baselines. Each query also includes a destination, in line with the baselines' preferences. The results are presented in Fig.~\ref{expfig:comparion-unsequenced-mu} -\ref{expfig:comparion-unsequenced-k}. This evaluation shows that KATR-Search outperforms ROSE-GM and OSSCaling significantly regardless of the values of $m$ and $k$. Among the baselines, ROSE-GM sets a rating threshold to exclude POIs with ratings below this threshold, resulting in a smaller performance margin compared to KATR-Search than OSSCaling. As $m$ and $k$ increase, the advantage of KATR-Search becoming more pronounced thanks to its superior pruning efficiency.

Given that the KATR query does not require a predetermined destination, we compare KATR-Search with the baselines ND-ROSE-GM and ND-OSScaling, which eliminate this requirement. This evaluation also incorporates the varying ratings of POIs. Results shown in Fig.~\ref{expfig:unsequenced-no-destination-mu}
demonstrate that KATR-Search has a significant priority on the query time over baselines, further accentuated with more query keywords. 

{{\bf Varying distance budget.} Given KATR-Search does not impose a travel budget like OSSCaling to limit the search space, we compare their query costs varying distance budget $\Delta$ to examine their efficiencies as the search space expands. We set $\Delta$ from 1 to 5 times the length of the longest route among the top-$k$ routes identified by KATR-Search.  Fig.~\ref{expfig:distance-budget-delta} shows that OSSCaling's query cost increases with larger $\Delta$ values, as its greedy search continues expanding even after finding the top-$k$ routes, if unvisited km-POIs remain. Conversely, KATR-Search maintains consistent query times by halting the search once the top-$k$ routes are identified.}

\subsubsection{Comparison in Flexible Visiting Order}
Among the benchmarks, only DAPrune and OSSScaling support flexible POIs visiting order, so our comparison focuses on these two baselines here. As DAPrune does not consider diverse ratings of POIs, we assigned identical ratings to all POIs for this assessment. The results, shown in Fig.~\ref{expfig:global-datasets}-\ref{expfig:global-k}, reveal that KATR-Search requires only 20\% of the query time of DAPrune, the runner-up, on each dataset, and is superior to OSSCaling by an order of magnitude. Furthermore, as illustrated in Fig.~\ref{expfig:global-k}, the query time growth for KATR-Search is significantly slower than that of the other two baselines as $k$ increases, demonstrating KATR-Search's superior performance in handling queries with larger $k$ values.

\begin{figure}[htb!]
  \centering
  \captionsetup{font={scriptsize}}  
\begin{subfigure}{0.49\linewidth}
      \centering   
\includegraphics[width=\textwidth]{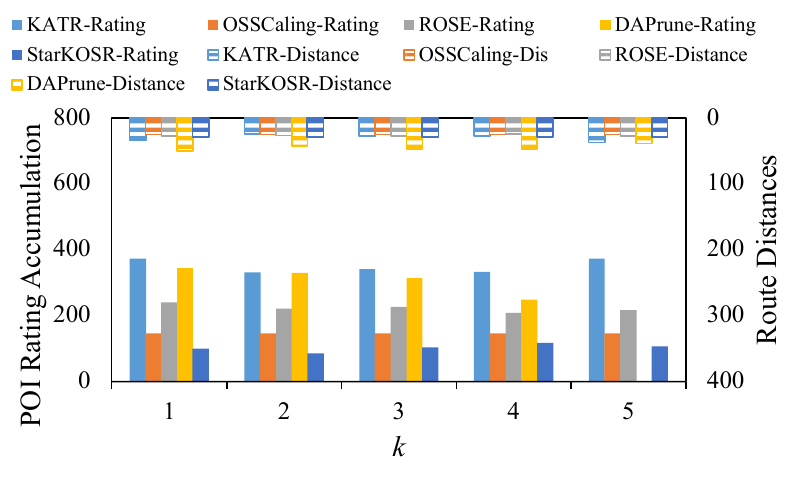}
    \vspace{-0.6cm}
    \caption{Route Effectiveness Comparison w.r.t. $k$}\label{effectiveness-comparison-k}
    \end{subfigure}      
\begin{subfigure}{0.48\linewidth}
      \centering   
\includegraphics[width=\textwidth]{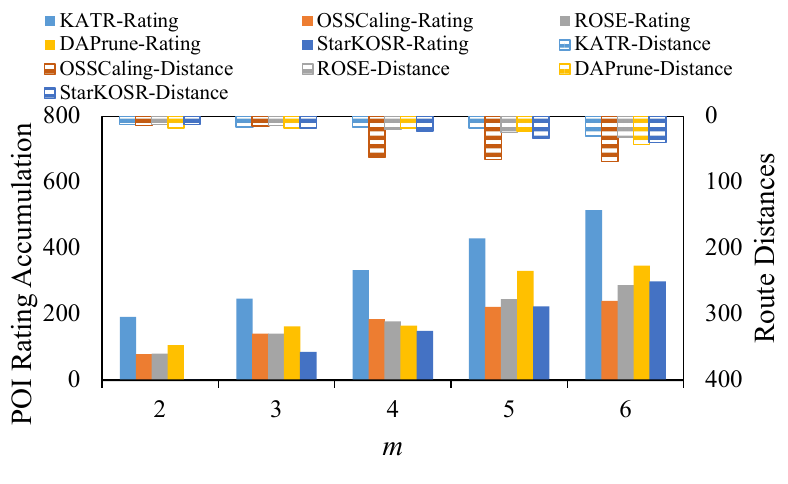}
    \vspace{-0.6cm}
    \caption{Route Effectiveness Comparison w.r.t. $m$}\label{effectiveness-comparison-m}
    \end{subfigure}  
\begin{subfigure}{0.49\linewidth}
      \centering   
\includegraphics[width=\textwidth]{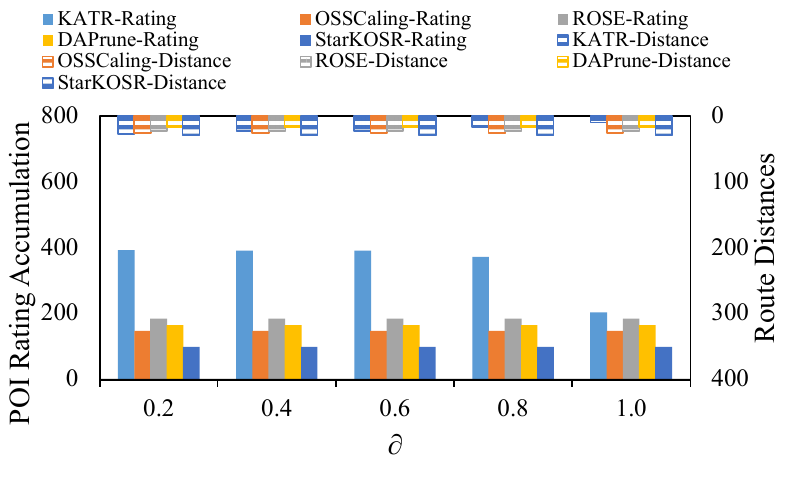}
    \vspace{-0.6cm}
    \caption{Route Effectiveness Comparison w.r.t. $\alpha$}\label{effectiveness-comparison-alpha}
    \end{subfigure}
    \begin{subfigure}{0.48\linewidth}
      \centering       
      \includegraphics[width=\textwidth]{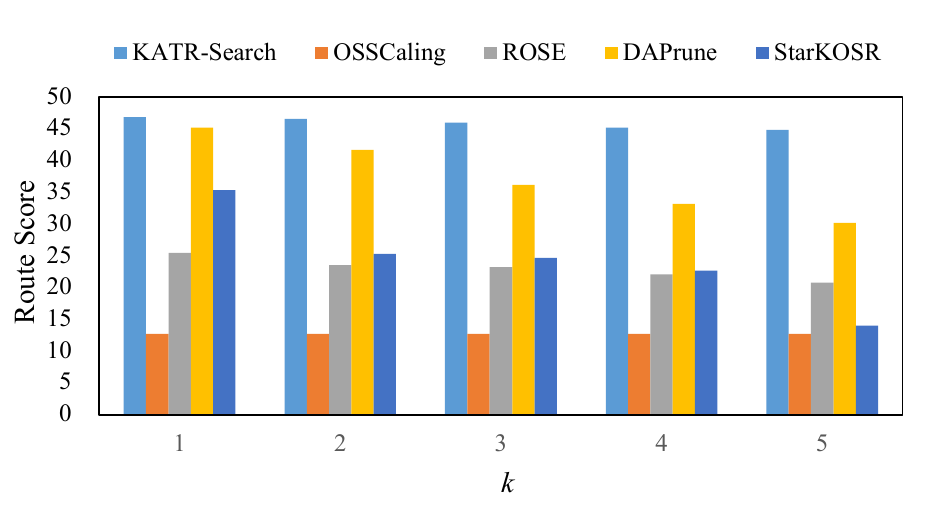}
    \vspace{-0.6cm}
    \caption{Route Score Comparison}\label{effectiveness-comparison-final-score}
    \end{subfigure}
\vspace{-0.2cm}
\caption{Route Effectiveness Comparison}
 \vspace{-0.4cm}
\end{figure}

{\subsubsection{Route Effectiveness Comparison}
 Finally, we compare the route quality of all approaches in terms of route distances, POI ratings, and final route scores. 
 Fig.~\ref{effectiveness-comparison-k} highlights the differences between KATR-Search and the baselines regarding route distances and POI rating accumulation, with fixed $m=4$ and $\alpha=0.8$, while varying $k$. The left $y$-axis displays the average accumulation of POI ratings along each route, and the right $y$-axis shows the average distance for each route, calculated as the sum of edge weights divided by 1,000. OSSCaling's performance remains unchanged with varying $k$ as it identifies only the optimal route. The results demonstrate that while the routes identified by KATR-Search may not always be the shortest, they consistently achieve the highest POI rating accumulation, leading to better final route scores, as seen in Fig.~\ref{effectiveness-comparison-final-score}. Similar trends are observed with varying $m$ and $\alpha$ as shown in Fig.~\ref{effectiveness-comparison-m}-\ref{effectiveness-comparison-alpha}, reinforcing our objective of allowing slightly longer routes to include higher-rated POIs.}  

\section{Conclusions}\label{sec:conclusion}
This work presents a flexible and comprehensive route planning strategy that can serve as an external tool for LLMs to provide convenient route planning services. It takes user-provided keywords and generates optimal routes passing relevant POIs, effectively accommodating various user preferences regarding route distance and POI ratings. To find top-$k$ routes, we developed KATR-Search, which progressively narrows the search scope by efficiently eliminating redundant candidates through a hierarchical pruning framework. This iterative process allows us to gradually uncover superior routes and refine the search space, ensuring the accurate identification of the top-$k$ routes as the search space is thoroughly explored.

  \section*{Acknowledgments}
This work was supported by NSFC Grants [No.62172351], ‌Shandong Provincial Natural Science Foundation [ZR2025MS1028], and NSERC Discovery Grants.

\bibliographystyle{IEEEtran}
\bibliography{reference}
\ifCLASSOPTIONcompsoc
\vspace{-0.3cm}
 
\vspace{-30pt}
\begin{IEEEbiography}[{\vspace{-30pt}\includegraphics[width=1in,height=1in,clip,keepaspectratio]{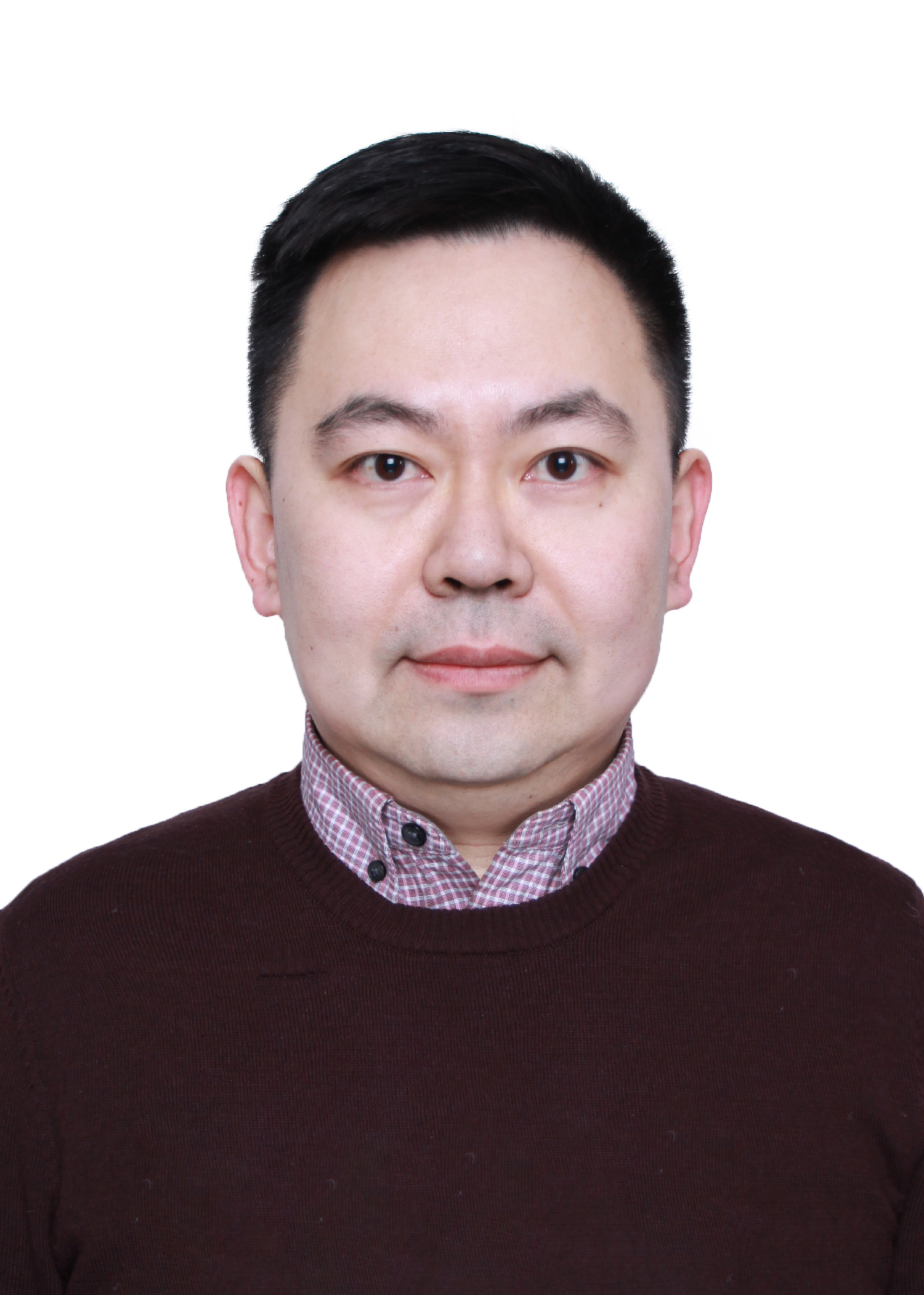}}]
{Ziqiang Yu} received his Ph.D. degree in computer science in 2015 from Shandong University, China. He is currently a professor in the School of Computer and Control Engineering, Yantai University, China. His research interests mainly focus on spatial-temporal data processing and graph computing. 
\end{IEEEbiography}

\vspace{-50pt}
\begin{IEEEbiography}[{\vspace{-40pt}\includegraphics[width=1in,height=1in,clip,keepaspectratio]{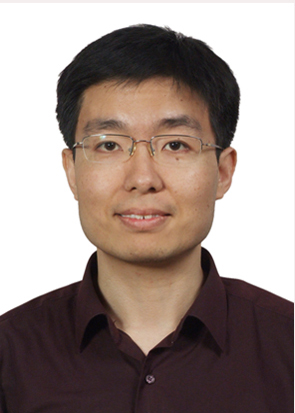}}]{Xiaohui Yu} received his Ph.D. degree in computer science in 2006 from the University of Toronto, Canada. He is currently a professor in the School of Information Technology, York University, Toronto.  His research interests are in the areas of database systems and data mining.
\end{IEEEbiography}

\vspace{-50pt}
\begin{IEEEbiography}[{\vspace{-40pt}\includegraphics[width=1in,height=1in,clip,keepaspectratio]{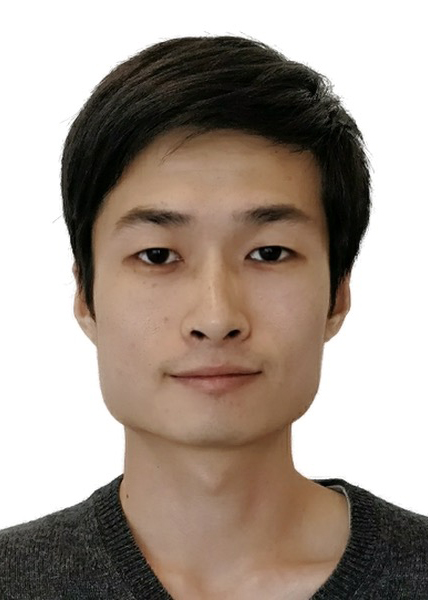}}]{Yueting Chen} received his Ph.D. degree in computer science from York University in 2023, supervised by Xiaohui Yu. He has worked as a postdoctoral fellow at York University and is currently a lecturer at Seattle University. His research interests mainly focus on data management systems.
\end{IEEEbiography}

\vspace{-50pt}
\begin{IEEEbiography}[{\vspace{-40pt}\includegraphics[width=1.1in,height=1in,clip,keepaspectratio]{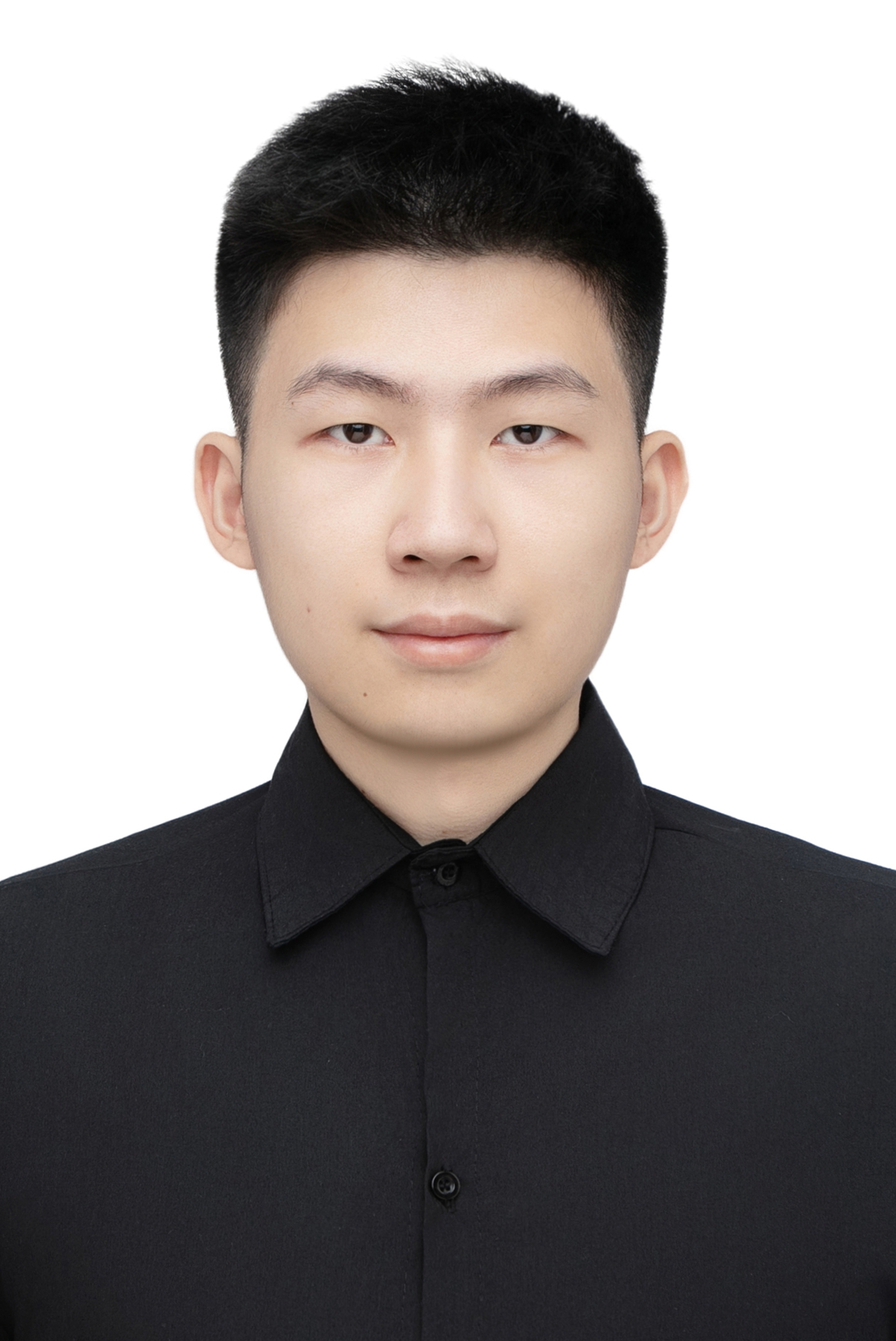}}]{Wei Liu} received his bachelor degree from Jinan University. He is currently pursuing a master's degree in the School of Computer and Control Engineering, Yantai University, China. His main areas of research are spatial-temporal data processing and graph computing.
\end{IEEEbiography}

\vspace{-50pt}
\begin{IEEEbiography}[{\vspace{-30pt}\includegraphics[width=1in,height=1in,clip,keepaspectratio]{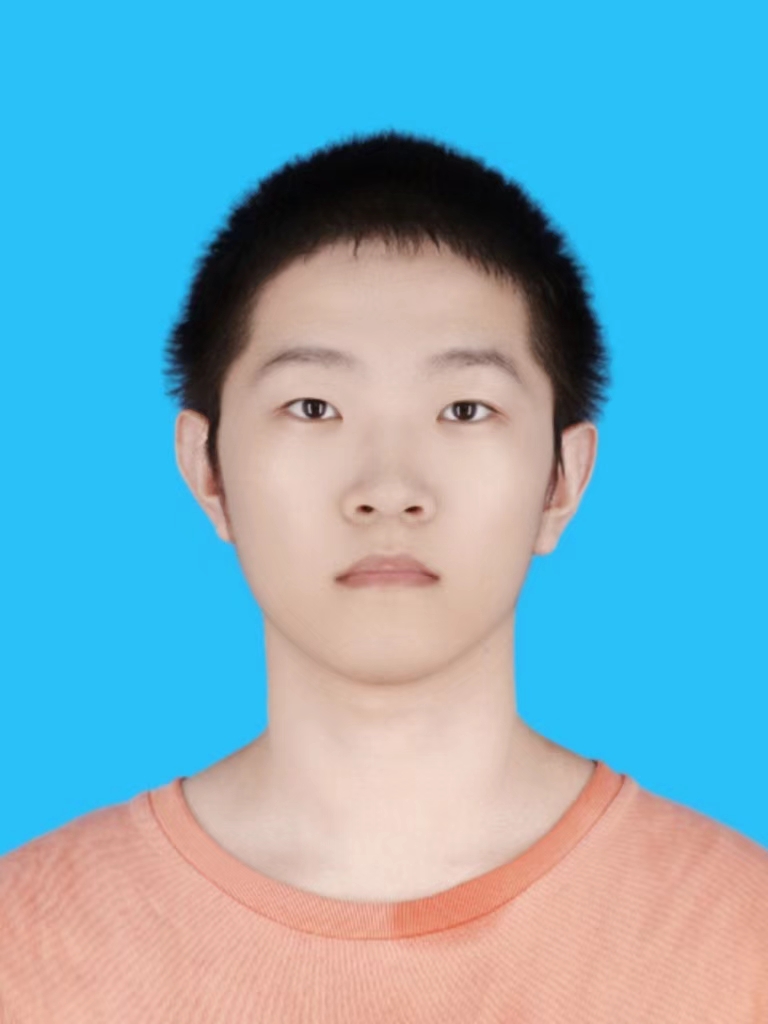}}]{Anbang Song}
received his bachelor's degree from Shanghai Institute of Technology. He is currently pursuing a master's degree in the School of Computer and Control Engineering, Yantai University, China. His main areas of research are spatial-temporal data processing and graph computing. 
\end{IEEEbiography}

\vspace{-50pt}
\begin{IEEEbiography}[{\vspace{-30pt}\includegraphics[width=1in,height=1in,clip,keepaspectratio]{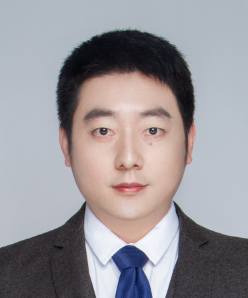}}]{Bolong Zheng} received his Ph.D. degree from the University of Queensland, Australia. He is currently a professor in  Wuhan University of Technology, China, where he also serves as the dean of the School of Computer Science and Artificial Intelligence. His research focuses on the data management and analysis.
\end{IEEEbiography}

\ifCLASSOPTIONcaptionsoff
  \newpage
\fi

\end{document}